\pdfoutput=1

\documentclass[5p,twocolumn,preprint,a4paper,12pt,times]{elsarticle}

\usepackage{amssymb}
\usepackage{wasysym}   
\usepackage{hyperref}

\usepackage{lineno}

\begin{document}

\begin{frontmatter}

\title{New complex EAS installation of the Tien~Shan Mountain\\Cosmic Ray Station}



\author[a]{A.P.Chubenko}

\author[a]{A.L.Shepetov}
\ead{ashep@www.tien-shan.org}

\author[b]{V.P.Antonova}

\author[a]{R.U.Beisembayev}


\author[a]{A.S.Borisov}

\author[a]{O.D.Dalkarov}

\author[b]{O.N.Kryakunova}

\author[e]{K.M.Mukashev}

\author[a]{R.A.Mukhamedshin}

\author[a]{R.A.Nam}

\author[b]{N.F.Nikolaevsky}

\author[a]{V.P.Pavlyuchenko}

\author[a]{V.V.Piscal}

\author[a]{V.S.Puchkov}

\author[a]{V.A.Ryabov}

\author[c]{T.Kh.Sadykov}

\author[d]{N.O.Saduev}

\author[b]{N.M.Salikhov}

\author[a]{S.B.Shaulov}

\author[a]{A.V.Stepanov}

\author[a]{N.G.Vildanov}

\author[a]{L.I.Vildanova}

\author[a]{M.I.Vildanova}

\author[c]{N.N.Zastrozhnova}

\author[a]{V.V.Zhukov}

\address[a]{P.N.Lebedev Physical Institute of the Russian Academy of Sciences (LPI), Leninsky~pr., 53, Moscow, Russia}
\address[b]{Institute of Ionosphere, Kamenskoye~plato, Almaty, Kazakhstan}
\address[c]{Institute for Physics and Technology, Ibragimova~str., 11, Almaty, Kazakhstan}
\address[d]{Institute for Experimental and Theoretical Physics of al-Farabi Kazakh National University, al-Farabi~pr., 71, Almaty, Kazakhstan}
\address[e]{Abay Kazakh National Pedagogic University, Dostyk~pr., 13, Almaty, Kazakhstan}

\begin{abstract}
In this paper we present a description of the new complex installation for the study of extensive air showers which was created at the Tien Shan mountain cosmic ray station, as well as the results of the test measurements made there in 2014-2016. At present, the system for registration of electromagnetic shower component consists of $\sim$100 detector points built on the basis of plastic scintillator plates with the sensitive area of 0.25m$^2$ and 1m$^2$, spread equidistantly over $\sim$10$^4$~m$^2$ space. The dynamic range of scintillation amplitude measurements is currently about $(3-7)\cdot 10^4$, and there is a prospect of it being extended up to $\sim$10$^6$.  The direction of shower arrival is defined by signal delays from a number of the  scintillators placed cross-wise at the periphery of the detector system. For the investigation of nuclear active shower components there was created a multi-tier 55m$^2$ ionization-neutron calorimeter with a sum absorber thickness of $\sim$1000g/cm$^2$, typical spatial resolution of the order of 10cm, and dynamic range of ionization measurement channel about $\sim$10$^5$. Also, the use of saturation-free neutron detectors is anticipated for registration of the high- and low-energy hadron components in the region of shower core. A complex of underground detectors is designed for the study of muonic and penetrative nuclear-active components of the shower.

The full stack of data acquisition, detector calibration, and shower parameters restoration procedures are now completed, and the newly obtained shower size spectrum and lateral distribution of shower particles occur in agreement with conventional data. Future studies in the field of $10^{14}-10^{17}$eV cosmic ray physics to be held at the new shower installation are discussed.
\end{abstract}


\end{frontmatter}


\section{Introduction}

\paragraph{The primary cosmic ray spectrum}
For more than five decades, the nature and origin of cosmic rays (CR) remain one of the main open issues in astrophysics, and this uncertainty does steadily increase with the growth of CR primary energy. The measurement of the cosmic rays energy spectrum, estimations of their composition, and the search for anisotropy of arrival directions have been the subject of many experiments.

\begin{figure}
\begin{center}
\includegraphics[width=0.5\textwidth, clip,trim=18mm 27mm 15mm 60mm] {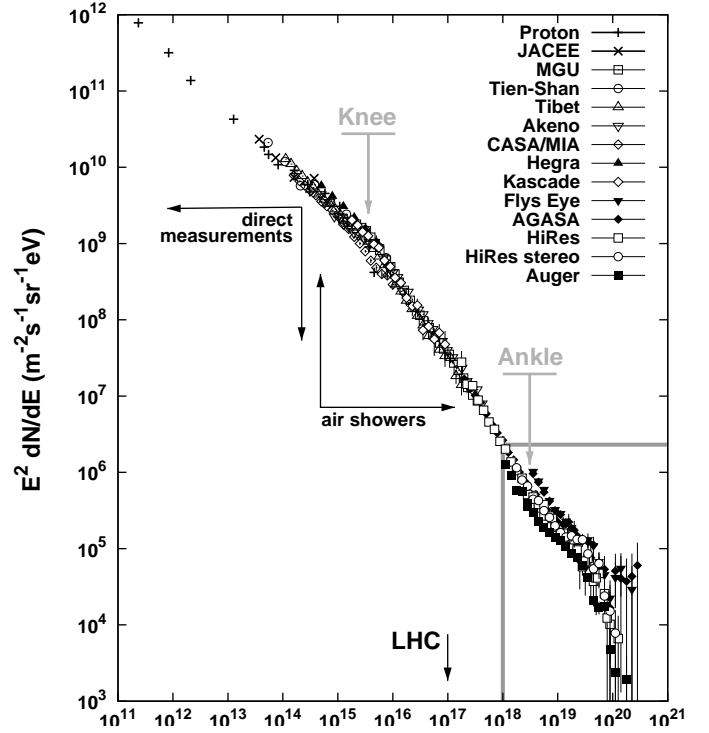}
\end{center}
\caption {The differential energy spectrum of the primary cosmic ray particles (the plot taken from publication \cite{2011letessier}).}
\label{fig0spc}
\end{figure}

As it can be seen in figure~\ref{fig0spc}, which is taken from the work \cite{2011letessier}, the intensity of differential energy spectrum of cosmic ray particles varies through 28~orders of magnitude over more than 10~decades of primary energy $E_0$. The  spectrum has a striking power-law behavior, $dN/dE \sim E^{-\gamma}$, with its power index $\gamma$ having several characteristic irregularities: around the energy $3\cdot 10^{15}$eV where the well known {\it knee} resides \cite{2005kascade}, a less prominent {\it second knee} approximately at $2\cdot 10^{17}$eV\cite{2001hires}, the {\it ankle} near $3\cdot 10^{18}$eV \cite{2010auger}, and the strong suppression around $5\cdot 10^{19}$eV \cite{2010auger_b},\cite{2015telescopearray}. Beyond the energy about $10^{10}$eV and until the knee the differential power law index $\gamma$ is about 2.7, above the first knee $\gamma$ is $\sim$3.0, with a small increase up to $\sim$3.3 at the second knee. Above the ankle, the spectrum is well described with $\gamma \sim$~2.7 again up to its cut-off near $5\cdot 10^{19}$eV, where the index roughly changes to $\gamma \sim$~4.3. Except the well-known Greisen–-Zatsepin-Kuzmin (GZK) effect \cite{1966gzk_a,1966gzk_b} which reveals itself as an abrupt spectrum cut-off at extremely high energies, the origin of all these features is still not clear enough, but they might arise from interference among different factors such as the loss of acceleration efficiency in galactic sources, the predominance of extra-galactic CR component at very high energies, and the impact of the CR propagation effects through interstellar medium. Therefore, in order to verify various scenarios for the origin of spectrum features it is necessary to perform accurate high-statistics measurements between $10^{15}$ and $10^{18}$eV.

Cosmic rays with the energy $E_0$ up to $10^{14}$eV can be studied directly through detection of primary particles by means of balloon
and satellite
based apparatus, but as the energy increases, the CR flux becomes too negligible for direct measurements, since it is impossible to use any detectors large enough for this purpose on the balloons or in space. On the other hand, the interaction cross-section of primary particles with $E_0\gtrsim 10^{14}$eV is quite sufficient for initiation of an atmospheric cascade due to succeeding collisions of CR particles with air nuclei, and numerous, wide spread components of the particle shower (called the extensive air shower --- EAS) can be detected at the ground \cite{oneasbook}. At those energies, the cosmic rays are investigated through observation of EAS with the use of ground based detectors which have the appropriate sum sensitive area, and which are placed on suitable altitude. The necessary aperture of any shower installation depends on the range of primary cosmic ray energies being under investigation. The CR particles which energy exceeds $10^{15}$eV can be effectively detected only with a shower array of the area above $10^4$~m$^2$.

Until now, investigations of the primary CR energy spectrum and composition in the energy range $E_0\lesssim 10^{17}$eV have been made with the help of a number of ground-based installations spread throughout the globe. Between them, the distribution of the charged EAS particles was studied with the use of different types of electronic particle detectors both at mountain sites such as Akeno \cite{1984akeno}, GAMMA \cite{2008aragats}, Tibet \cite{2008tibet}, and at the sea level by KASCADE \cite{2005kascade} and KASCADE-Grand \cite{2012kascade} experiments.  The Cherenkov and fluorescent light emitted by EAS particles was registered by the Fly's~Eye \cite{1994flyseye}, HEGRA \cite{2000hegra}, Tunka \cite{2012tunka}, and Yakutsk \cite{2015yakutsk} arrays. The nuclear-active component of EAS was studied by the Tien~Shan ionization calorimeter \cite{scinti1}, and the properties of high-energy hadronic interactions in the mountain experiments Pamir \cite{xrec_trudy154} and Chacaltaya \cite{2008chacaltaya} which were based on the method of X-ray emulsion chambers (XREC). In the complex Hadron experiment at Tien~Shan the combination of EAS and XREC methods \cite{ontienhistory} was applied.

The installations mentioned here have registered a multitude of EAS components at different scales of energy and spatial variation :
\begin{itemize}
{\item shower electrons and positrons with the energy $E_e\gtrsim 1$MeV, multiplicity $N_e \sim10^5-10^9$, and typical lateral distribution within a $R_e\lesssim1$km range;}
{\item the Cherenkov radiation of relativistic shower particles at a distance up to 100m from EAS core;}
{\item low-energy muons with $E_\mu\gtrsim 1$GeV, $N_\mu \sim 10^2-10^6$, and $R_\mu\lesssim100$m;}
{\item high-energy muons with $E_\mu\gtrsim 200$GeV, $N_\mu \sim 10-1000$, and $R_\mu\lesssim1$m;}
{\item the flow of hadronic (nuclear active) particles with $E_h\gtrsim 1$GeV in a close vicinity of EAS core, $R_h\lesssim 10$m;}
{\item genetically connected groups ({\it families}) of high-energy hadrons and "gamma-quanta" ($\gamma$, $e^-$, $e^+$) with the energy above $2-4$TeV and typical lateral size of the order of $2-10$cm registered within the EAS core region by XREC method;}
{\item evaporation neutrons in a wide energy range, from thermal energies of the order of $E_n\sim10^{-2}$eV, and up to some tens and hundreds of GeV, both around the EAS core and at its periphery up to some tens of meters;}
{\item $30-300$MHz radio-emission from EAS particles.}
\end{itemize}
All these characteristics are sensitive both to elementary composition of primary cosmic rays and fundamental properties of strong interaction, though in different proportions; hence, the detection of any of them is complimentary to each other. Ideally, any array intended for EAS study should detect simultaneously all the components but the majority of real experiments have registered only one or few of them. Besides, a large part of existing and past shower installations do reside at the sea level where the EAS observation is less appropriate than at the high mountain altitudes for a number of reasons.

Investigation of the $10^{14}-10^{17}$eV EAS with a relatively compact detector system is mostly effective at the mountain height ($3000-4000$m above the sea level, and with integral atmosphere thickness around $600-700$g/cm$^2$), since at these altitudes the showers occur being close to the maximum of their development, reaching the highest value and minimal fluctuation of particle multiplicity. Besides, this observation level is nearer to the point of primary interaction, so the reconstruction of shower parameters becomes more reliable, and their values are more sensitive to predictions given by the particle interaction models to be verified. The {\it Mollier radius} which defines the lateral scattering of EAS particles is somewhat larger at high altitude, and correspondingly the spatial distribution of shower particles is flatter resulting in a smaller probability to miss the shower than at the sea level. These specific features of mountain installations give a unique possibility to study the properties of primary cosmic ray  particles, both from the nuclear physics and astrophysics points of view, and in particular to elucidate the nature of the knee in primary CR spectrum which remains one of the most urgent problems in the cosmic ray physics since its discovery in 1958.

Most of the CR installations listed above were aimed primarily for investigation of high-energy cosmic rays in the energy range of $E_0\gtrsim 10^{15}-10^{18}$eV, so the knee region occurred just at the beginning of its operation diapason. With exception of XREC experiments, this orientation forced to use a rather spaced distribution of particle detectors, with a typical spatial step between the neighbouring detector points of the order of tens of meters to overlap the sum installation area necessary for registration of low intensity events at the upper end of the energy range of interest. With such a geometry of detector disposition the gaps between them occur of the same order, or even larger than a typical size of the EAS core ($\lesssim 1-3$m) which makes impossible any precise study of the latter. For the same reason, the modern gigantic arrays destined for the study of extremely high-energy end of the CR spectrum such as Auger \cite{2001auger}, IceTop \cite{2008icetop}, and Telescope Array \cite{2012telescopearray} also occur inappropriate for registration of the particle distribution within the central part of extensive air shower.

At the same time, it is just the core region of EAS which is crucially important to solve the problem of the knee in primary cosmic ray spectrum.

\section{The cosmic ray study at the Tien~Shan mountain station}

\paragraph{Investigations in the EAS core region}

Traditionally, the study of the central part of extensive air shower was one of the leading problems which were solved at the Tien~Shan mountain cosmic ray station where successive generations of physical set-ups have been used for this purpose: the deep ($36$m$^2$) lead calorimeter \cite{scinti1}, Wilson chamber \cite{ontienvilson}, spark chamber \cite{ontiensparkchamber}, and the large-size ($162$m$^2$) X-ray emulsion chamber \cite{xrec_trudy154}. Because of its $\sim 0.1$mm spatial resolution and the possibility to measure the energy of separate hadrons and gamma-quanta the latter method is most adequate for the purpose of center EAS study but it also has a significant disadvantage: strong fluctuations of EAS core development in the atmosphere present the obstacle for determination of the primary energy $E_0$ of registered events. In the hybrid Hadron \cite{ontienhistory} installation this problem was solved through combination of the X-ray emulsion technique with electronic detectors in a joint XREC+EAS set-up.

Besides determination of the energy spectrum of primary cosmic rays and energy dependence of different EAS characteristics in the knee region, as a result of these measurements some uncommon effects where found which can hardly be in agreement, and some of them do explicitly contradict to existing standard models of shower development in the atmosphere, such the anomalous increase of the absorption length of shower cascade at the primary energy $E_0\gtrsim 10^{15}$eV (the so called effect of long-flying component) \cite{longflying2,longflying3,longflying_pamir,longflying_pamir2}; the gamma-families with {\it halo} \cite{halo1,halo2,halo3}, the alignment of the most energetic cascades in the gamma- and hadron families \cite{alignment1,alignment2,alignment3}, anomalously high muon multiplicity in the EAS with gamma-families \cite{hadron_high_muons}, abrupt scaling violation in the energy spectra of particular gamma-quanta in gamma-families \cite{hadron_scaling}. A common characteristic feature of all these effects is their tendency to group just within the central part of EAS where the most energetic hadrons are concentrated which conserve the movement direction of the primary cosmic ray particle in the laboratory coordinate frame. Later on, in the experiments with neutron detectors operating synchronously with EAS installation the effect of superfluous production of evaporation neutrons was revealed both in surface \cite{jopg2001,jopg2008} and underground neutron detectors \cite{undgour1,undgour2} which is also typical for the core region of EAS.

Furthermore, some anomalous CR interaction events have been observed in XREC experiment at Chacaltaya mountain which have an abnormal multiplicity ratio of their charged and neutral hadronic products \cite{centauro0}; later on, similar events (named {\it Centauros} and {\it Anti-Centauros}) were discovered by the Pamir experiment \cite{centauro2} and at balloons \cite{centauro1}. The isospin symmetry which is commonly kept in strong interaction of usual hadrons is roughly violated in the events of such type, so that there has been made a hypothesis that in the primary CR flow there is some abnormal component constituted by the particles of the strange quark matter \cite{centauro_bjorken}. In \cite{centauro_sqm} it is shown that at least some of anomalous CR effects seen in high-altitude experiments at Chacaltaya, Pamir, and Tien~Shan could be explained by the presence of the strange matter particles (called {\it strangelets}).

However, generally in modern collider experiments it has not yet been found any principal deviation from the predictions of the Standard Model in the range of center-of-mass energies equivalent to  $E_0\sim 10^{15}-10^{16}$eV where the mentioned anomalous effects are just met quite prominently in cosmic ray interactions. This contradiction may occur a rather illusive one if to take into account that the typical geometry in the case of a EAS core region in a CR experiment with its characteristic size of the order of $\lesssim5-10$m and the altitude of the primary particle interaction about $20-30$km above observation level corresponds to the pseudorapidity range about $\eta \sim$12, which is practically unseen at any collider because of specific constructive restriction imposed by detector positioning around vacuum pipes. Moreover, the natural flux of high energy cosmic rays may probably contain some unusual particles, such as strangelets, which could give a rise to abnormal phenomena observed in CR but are absent or undetectable at accelerator. Thus, cosmic ray experiments can provide a complementary data channel in the study of high energy hadronic interactions which remain principally inaccessible with accelerator technique.

To summarize, one can state that up to the present time some phenomena have been observed in various cosmic ray experiments which still remain unclear at current stage of our knowledge on particle interactions, and the modern simulations of EAS development in atmosphere cannot adequately explain the massive of known experimental results as a whole. Also, a number of open questions connected with the nuclear physics aspect of CR study do exist which can not be completely settled by the modern accelerator-based methods. These problems demand a precise investigation of the properties of different EAS components, in particular in the vicinity of shower center since this is the region where the most energetic EAS hadrons are concentrated, and there is quite evident tendency towards near-core concentration of all anomalous CR effects as well. To trace precisely the energy dependence of different EAS characteristics trough their passage over the knee region of primary spectrum can be helpful to understand the origin of the knee  if it is connected with appearance of some unusual CR components, or with some change in the elementary process of nuclear interaction, or with some local peculiarity of the CR acceleration mechanism. For such a purpose, a rather dense, preferably continuous disposition of different kinds of particle detectors within the compact central part of cosmic ray installation, a sufficiently large aperture of this center, and a wide dynamic diapason of particle density registration are highly desirable. It is also essentially necessary to install the CR detector complex on a high-altitude mountain site.

\paragraph{The distribution of the EAS arrival directions}
The astrophysical aspect of the knee nature is connected with acceleration process of cosmic ray particles, and with their further propagation in Galaxy. Up to the present time, none of the multiple explanation models of the knee origin has yet obtained any reliable experimental confirmation. To the main reason for it is the distortion of the trajectories of charged CR particles by regular and chaotic magnetic fields on their way from the CR source to the Earth. Multiple scattering tangles particles' trajectories and blur the deposit of different CR sources on the sky so the local anisotropy of CR arrival directions does not exceed 1\% at the knee energies. On a large scale, propagation of cosmic rays should be considered as a diffusion process.

Nevertheless, if we trace the CR mass {\it composition} instead of their local intensity at different arrival directions there arises a possibility to get rid of the interference caused by magnetic scattering, and to obtain an anisotropy with sufficient statistics \cite{pavlyuchenko,pavlyuchenko2}. Moreover, the multiple scattering permits to inspect a great portion of the whole celestial sphere even with an installation which has only a limited coverage sector in equatorial coordinate frame. It is obvious that this kind of investigation demands the use of angular sensitive detector system for registration of EAS particles. Traditionally, the measurements of the pulse front delays between spaced particle detectors can be applied for this purpose with further restoration of EAS directional angles by fitting experimental delay distribution through minimization of a $\chi^2$ type functional.

\paragraph{The physics of cosmic ray muons}
Of a particular importance for understanding many unsolved problems in high-energy cosmic ray physics is investigation of the muonic component of EAS which circumstance is due to penetrative properties of muon particles.

When a very high energy primary CR particle interacts somewhere in the upper atmosphere, a multitude of secondary products, mainly pions and kaons are produced which determine the whole future development of shower cascade in the air. The secondary mesons either interact in their turn and produce the next generation of hadronic cascades with lower energy, or decay into high-energy muons which can be detected directly at ground level, or even deep underground. Both energy losses and angular deflection of muons on their way to surface detector array are negligibly small, so they provide a unique channel for the immediate study of the high-energy hadronic interaction at the very beginning of the particle cascade.

At the ground level, the muonic component of EAS depends on a number of succeeding particle coincidences which take place at all stages of the air shower cascade, and its properties occur being sensitive to many characteristics of the elementary process of high-energy hadron interaction, such as multiplicity, elasticity, charge ratio, and probability of the baryon/anti-baryon pair production \cite{muons1,muons2}. According to Monte-Carlo simulations of an EAS development in atmosphere, the total muon multiplicity occurs strongly depending on the parameters of adopted interaction model \cite{muons3,muons4}.

Besides the restriction put on a range of admissible interaction models, the investigation of the muon EAS component is potentially capable to minimize the uncertainty of the primary mass composition. Indeed, the muon abundance of the air shower induced by a nucleus consisting of $A$ nucleons is approximately $A^{0.1}$ times higher than that in the proton induced shower of the same energy; e.g. an EAS originating from interaction of an iron nucleus has by 40\% more muons than a proton one. Such dependence of the sum EAS muon number N$_{\mu}$ on the mass of primary CR particle is complementary to another mass-sensitive method of shower exploration which is based on determination of the depth of the shower maximum $X_{max}$ in atmosphere. By combination of both methods theoretical uncertainty in selection of possible hadron interaction models can be further constrained.

Of a particular interest is the observation of inclined extensive air showers which come with a large zenith angle, $\theta \gtrsim 60^o$. It is the high-energy muon component which dominates in such near-horizontal EAS since the most part of electromagnetic particles in this case is effectively absorbed due to a large distance passed by the shower on its way through the atmosphere. Therefore, the study of the strongly inclined showers provides a direct measurement channel of their muonic number N$_{\mu}$ by a ground-based detector system \cite{muons-horizo,muons-horizo2}.

Experiments on registration of the muonic EAS component have remained a traditional direction of CR investigations at the Tien~Shan mountain station \cite{ontienmuons1,ontienmuons2,ontienmuons3} where the whole infrastructure necessary for this purpose (a subsurface tunnel and a number of large area underground rooms) is present since the very beginning of its existence.

\paragraph{The Tien~Shan detector array and the goals of its modernization}
Generally, there exists a need to advance onto a qualitatively new investigation level in contemporary CR physics research. A new generation of an EAS detector array for effective investigation of the $E_0\sim 10^{14}-10^{17}$eV energy range should give a precise information on maximally complete set of shower characteristics including its electron-photon, hadronic, and muonic components. It should be possible to carry out a saturation-free measurement of a rather dense particle flow immediately in the core region of an EAS. For precise investigation of electron, hadron, and muon components around the high-energy EAS core the installation must have a large-sized (of some tens of square meters) continuous central detector part with an enhanced spatial resolution (of at least $10-20$cm, or better), and sufficient dynamic diapason of particle density measurements (up to $(1-2)\cdot 10^6$). At the same time, the wide-spread peripheral part of shower installation (at least, $10^4-10^5$m$^2$) is necessary to give sufficient data set for restoration of basic EAS characteristics: the position of shower axis in the plane of detector system with accuracy of $0.5-1$m or better, the shower {\it size} and {\it age},  energy estimation of the primary CR particle, and the direction angles of its arrival. Mountain high-altitude disposition ($3000-5000$m above the sea level) of the whole detector array is highly desirable.

In accordance with this challenge, the complex installation for EAS study which has existed at Tien~Shan mountain cosmic ray station since the middle of 1960s \cite{scinti1, ontienhistory,  ontienhistory3} demanded considerable modernization, including essential increase of the total area of detector covering, and upgrade of the most part of  the detector system by the new generation of particle detectors equipped with modern data acquisition electronics, so as the widening of the range of physical problems studied at Tien~Shan. Favorable high-altitude location and the planned complex perspective of the multipurpose detector set-up of the new Tien~Shan installation permit to enlarge considerably the comprehension of the collected experimental data, and to gain one of the best world data sets in the field of  cosmic ray physics and connected phenomena within the energy range around the knee of primary spectrum.

The aim of this paper is to give an overview of the current status of Tien~Shan detector complex, to describe the operation procedure anticipated for using in EAS data treatment, and to present physical results of a number of test measurements which have been made after the modernization of its main detector subsystems: two scintillation carpets for registration of the near-core distribution of EAS particles, ionization detector channels of hybrid calorimeter, and the neutron detectors of the nuclear-active EAS components.

\section{General layout of the Tien~Shan EAS installation}

\begin{figure*}
\begin{center}
\includegraphics[width=0.5\textwidth, clip,trim=10mm 40mm 7mm 40mm] {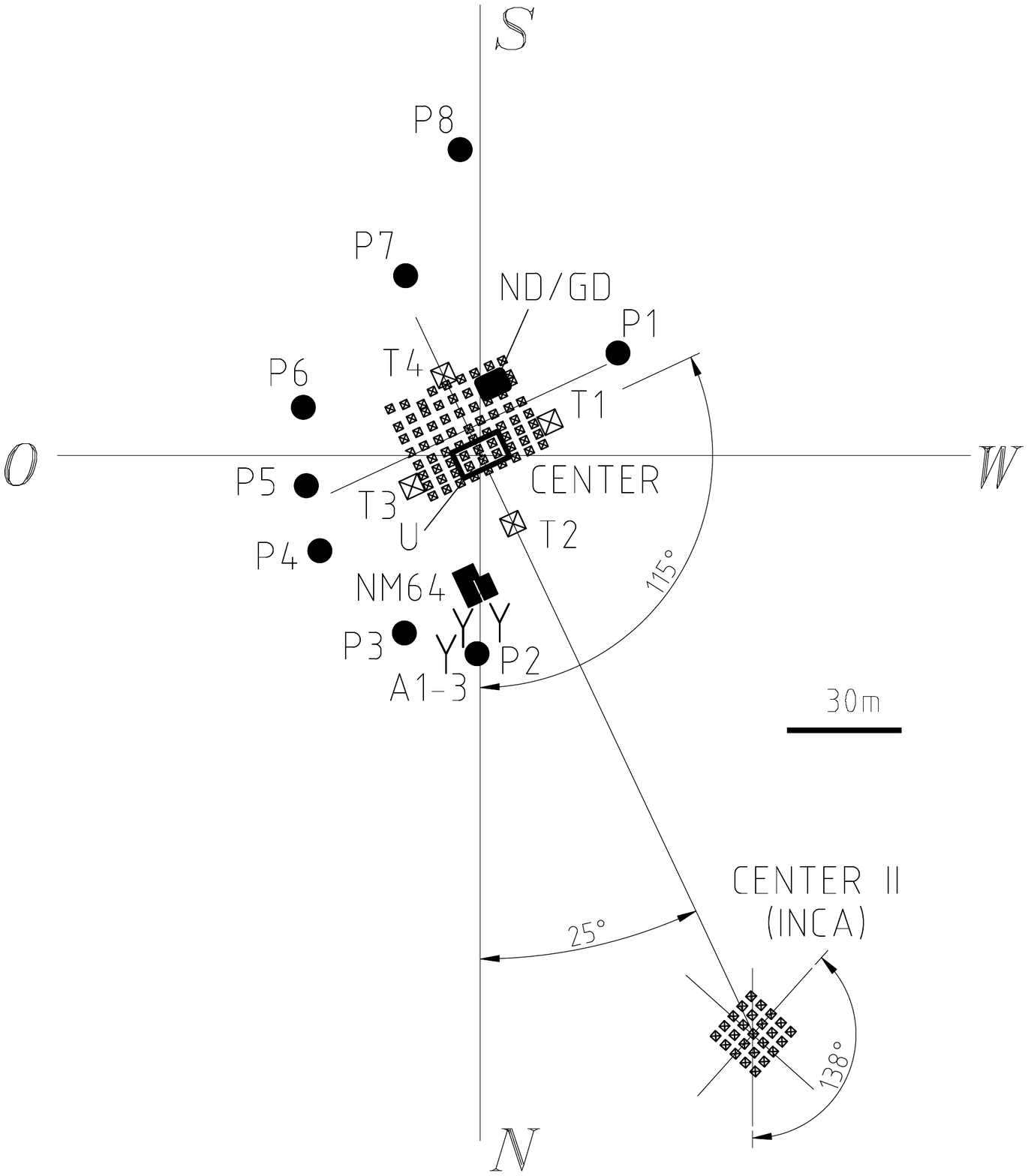}
\end{center}
\caption {General plot of the Tien~Shan EAS detector system.}
\label{figplan}
\end{figure*}

The Tien~Shan cosmic ray station is situated in the mountains of Northern Tien~Shan, 43$^\circ$~North, 75$^\circ$~East at the altitude of 3340m above the sea level. The average atmosphere depth at the height of station location is 690g/cm$^2$.

The disposition scheme of the new EAS detector complex on the territory of Tien~Shan station is presented in the figure~\ref{figplan}. Principally, the shower system is build around two points, {\it Center} and {\it Center~II},
where the shower core detectors are placed: the dense scintillation "carpet" for registration of the EAS electron component, a set of timing scintillators {\it T1-T4} for determination of the EAS arrival directions according to relative time delays of shower front, the complex of muon and hadron detectors in the underground room {\it U} situated beneath the {\it Center} point; another scintillation carpet, and the big ionization calorimeter IN\-CA in {\it Center~II}. There is also, the NM64 type neutron supermonitor placed in vicinity to the first detector center which is used as a detector of EAS hadron component. A number of additional scintillation detectors should be spread nearly concentrically over the station territory around the two center cores for registration of peripheral EAS particles.

The scintillation carpets of shower installation are designed for the study of the properties of EAS electromagnetic component in immediate vicinity to shower core. Accordingly to this purpose, both detectors and the whole subsequent tract of signal acquisition used in the carpets are planned to have an enhanced up to $(1-2)\cdot10^6$~dynamic range of scintillation amplitude measurement. The reason for a rather dense detector disposition is the intention to investigate possibly precisely the structure of lateral distribution of the particles in EAS core region. At present, there are 72~scintillation detectors installed in the 32$\times$27m$^2$ large {\it Center} carpet with basic spatial step 3$\times$4m, and 30~ones distributed with a step of 5$\times$6m over the $\sim$700m$^2$ area in {\it Center~II}. In 2016-2017~years it is planned also to arrange peripheral scintillation detectors at the distances of $20-200$m around both center carpets, and to connect them to free informational channels of the {\it Center} and {\it Center~II} registration systems.

Scintillation detectors of the each carpet subsystem are connected to its own trigger generation scheme which can provide a general control signal at the time when the momentary sum of scintillation amplitudes in the whole carpet is high enough to exceed some preset threshold. This signal is used as a trigger which marks the passage moment of an EAS, and initiates the sampling process of scintillation amplitudes and data registration in the corresponding data acquisition center. The detector points of {\it Center} and {\it Center~II} carpets can operate under the control of either its own local trigger or of an external one, which is transmitted from the neighbouring center via a cable line to ensure synchronous operation of both registration subsystems.

Output pulses from a group of scintillation detectors marked as {\it T1-T4} in figure~\ref{figplan}, which reside cross-wise around the {\it Center} carpet at the distances about 20m and 40m are connected to a digital oscilloscope with a 4ns resolution time scale for precise measurement of their mutual delays between the moments of EAS front arrival. These time delays are used in a $\chi^2$~minimization procedure which permits to define the primary direction of shower axis in horizontal and equatorial coordinate systems.

The underground room of the Tien~Shan station is located just below the {\it Center} detector carpet so the shower trigger signal of the latter can be conveniently applied for synchronization of the data recording process there. The depth of rock soil absorber above the room is 2000g/cm$^2$, which is equivalent to 5GeV energy threshold for penetrating muons. The complex of underground particle detectors includes the muon hodoscope on ionization counter tubes with the sum sensitive area about 80m$^2$, and two large-sized hadron detectors on the basis of neutron-sensitive ionization counters: the vertical neutron calorimeter and the horizontal neutron hodoscope. Simultaneously, the layers of heavy iron and lead substances which are a part of both neutron installations are used as an absorber filter in telescopic set-ups, based on coincidence signal from two opposite sets of scintillation detectors for selection of the high-energy muons traversing the underground room in the nearly vertical and horizontal directions. With addition of thick absorber layers within the neutron detectors the overall energy threshold for muon registration can be enhanced up to 7-8GeV.

The hybrid ionization-neutron calorimeter {\it IN\-CA} which resides in  {\it Center~II} point has an area of 55m$^2$ and a 1030g/cm$^2$ sum thickness. The main sensitive elements of calorimeter are the 300 $\times$ 11 $\times$ 6~cm$^3$ large ionization chambers of rectangular cross-sec\-ti\-on grouped into 9 registration layers. Long axes of the chambers in every pair of successive layers are directed perpendicularly to each other, so the 2D distribution of particle density can be obtained at every registration level, and traced step-by-step over the whole calorimeter depth. The use of ionization neutron counters for registration of the low-energy part of evaporation neutron spectrum, and of ionization gamma-ray counters is also anticipated in the IN\-CA construction. Above the upper top of calorimeter it is planned to install an X-ray emulsion film chamber.

The main advantage of the complex calorimetric setup in combination with scintillation shower carpet in {\it Center~II} point is a possibility to determine the energy of the primary cosmic radiation as well as a possibility to measure the angular, lateral, and longitudinal distributions of secondary particles both in the atmosphere and within the lead absorber. While the system of scintillation shower detectors and the ionization chambers inside the calorimeter have the spatial resolution of the order of $\pm$1m and $\pm$10cm correspondingly, the lateral resolution of X-ray chambers is as high as $\sim$100~microns. The accuracy of energy determination with X-ray method is also rather high, $\Delta E/E\sim 25$\%. However, every X-ray film accumulates a big number of events during the whole cycle of its exposition (about one year), and it is not an easy task to separate any genetically unrelated events at processing of experimental data. In spite of considerable complexity in operation, and the high cost of large-sized electronic detectors, such as scintillators and ionization chambers, they have a great advantage of a reliable time binding of every registered event. We believe that the most promising direction in development of the cosmic ray interaction studies is the hybrid one (X-ray chambers + EAS detectors) which combines the advantages of both techniques.

Complete experimental information obtained in operation of the shower detector system as a whole is stored in the common database of the Tien~Shan mountain station \cite{tieneng} for its further off-line analysis.

Current status of the Tien~Shan EAS detector complex is the following. All detectors in both scintillation carpets are set into their operation places and can give physical information, so the test measurement runs were held there in the years 2014-2016 (see later in this article). The fast timing measurements of scintillation delays in shower detectors for restoration of the EAS axis direction have been started in the end of 2015.  At IN\-CA calorimeter, four top ionization chamber layers are completely set into a ready-to-use state, and the test physical run is going on there. The installation of the chambers of the layers 5 and 6, and the tuning of their electronics are now in progress. At the underground complex, the detectors of the neutron calorimeter and muon hodoscope are mostly installed in their operation places together with all necessary electronics; the measurements on the neutron monitor surrounded with scintillation detectors of the muon telescope system have been constantly taken in the whole last decade.  The Tien~Shan NM64 neutron supermonitor also operates continuously during the years after a series of succeeding modifications of its electronic equipment.

\section{The scintillation shower system}
\label{sesci}
\begin{figure*}
\begin{center}
\includegraphics[width=0.49\textwidth, trim=0mm 60mm 0mm 70mm]{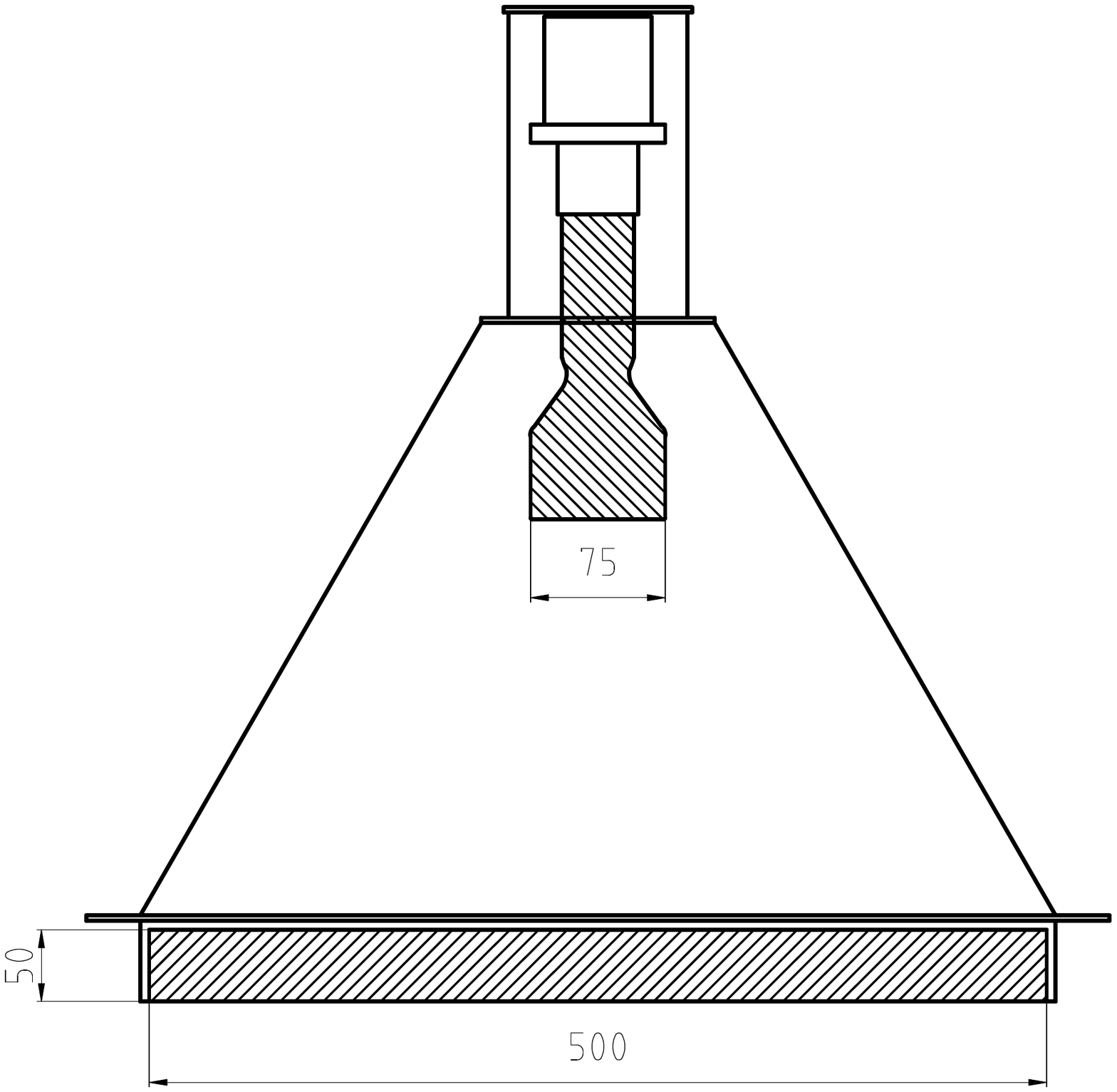}
\includegraphics[height=0.49\textwidth, angle=90, trim=0mm 20mm 0mm 0mm]{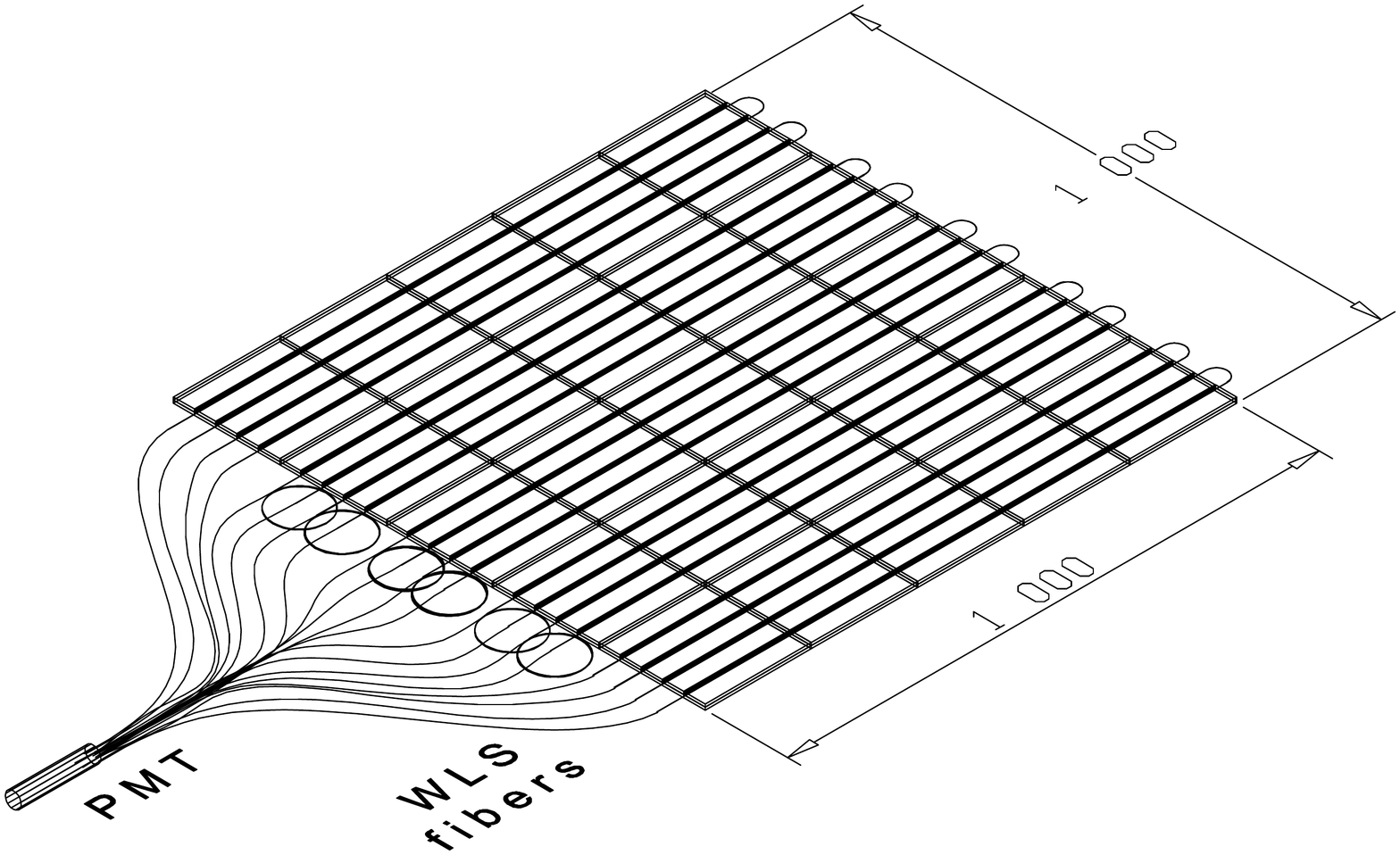}
\end{center}
\caption {Two types of scintillation detector used in the shower installation of the Tien~Shan station (dimensions are shown in mm).
}
\label{figscinti}
\end{figure*}

Nowadays, most part of the new shower system at Tien~Shan is built on the basis of particle detectors with a plastic 0.5$\times$0.5$\times$0.05m$^3$ scintillator and a pyramid-shaped light reflector which permits registration of the scattered scintillation light by a photomultiplier tube (PMT) \cite{scinti1}. Contemporary modification of this rather traditional detector type consists of full substitution of its internal electronic equipment with modern electronics: at present time, the build-in analog amplifiers based on the integrated circuit chips and super-high current transfer transistors are used for transmission of the PMT output pulses to the data registration center.

Another kind of particle detectors at Tien~Shan is based on the resent development of a large-sized 1$\times$1$\times$0.01m$^3$ plastic scintillator with the wavelength shifting fibers as a collection and output means of scintillation light \cite{shalour2006}. Due to their large sensitive area and advanced time resolution, detectors of this type are especially suitable for use in the time delay based measurement system of the EAS arrival angles.

The schematics of the internal arrangement of both detector types are presented in the figure~\ref{figscinti}.

\begin{figure*}
\begin{center}
\includegraphics[width=1\textwidth, clip,trim=10mm 55mm 10mm 55mm] {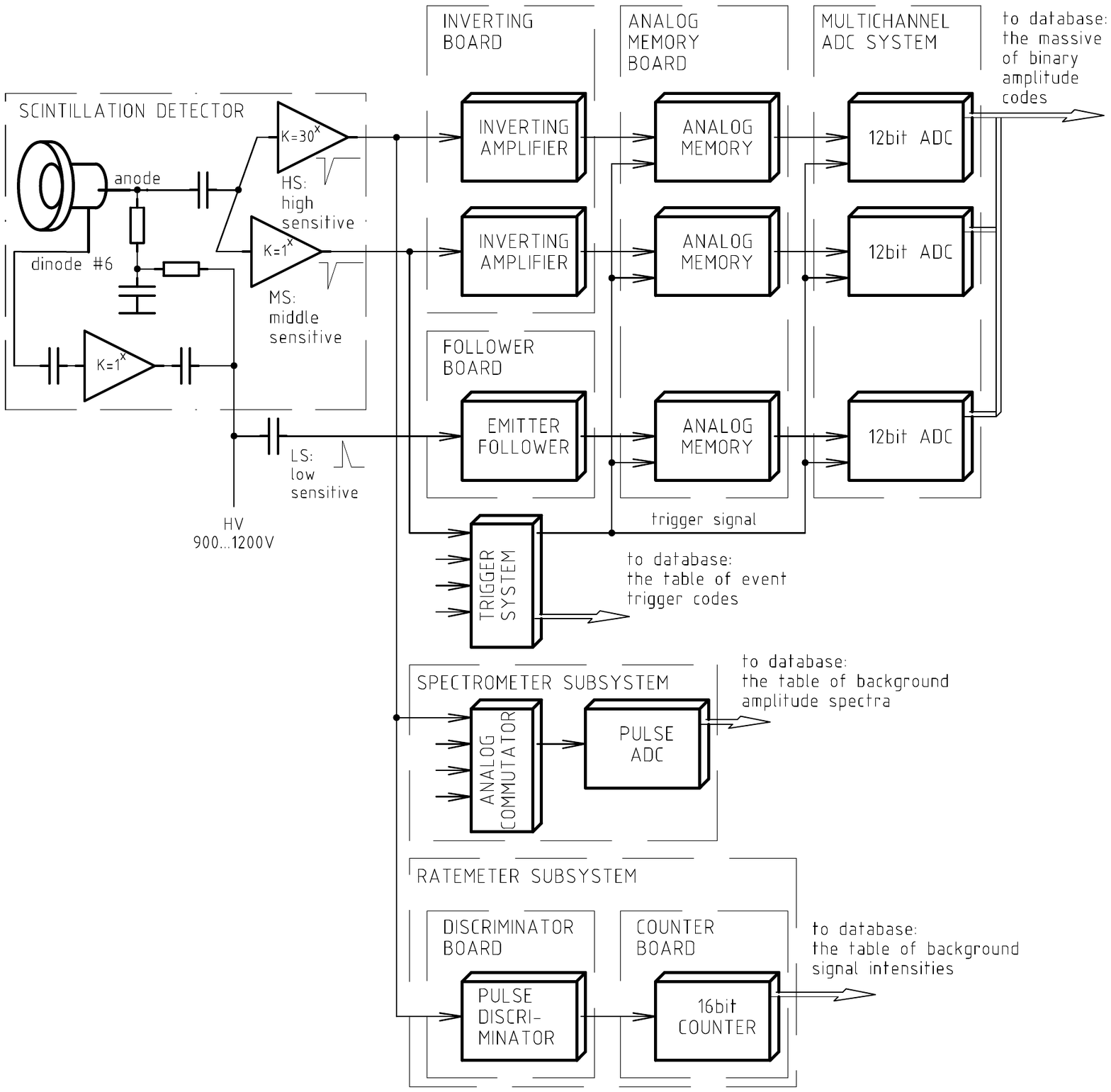}
\end{center}
\caption {Block diagram of the scintillation signal operation in a single detector channel of shower system.}
\label{figidaq}
\end{figure*}

The data acquisition procedure accepted in informational channels of scintillation shower system is illustrated by the figure~\ref{figidaq}. The scintillation pulse signals from the detectors distributed over the territory of Tien~Shan station are transmitted into the common data registration center over the long cable lines through the emitter follower schemes on powerful super-high current transfer transistors which reside immediately within the PMT casing. Each scintillation detector has just three output connectors where the analog pulse signals from its photomultiplier tube are switched to, providing simultaneously three different amplitude channels operating in parallel. To the high-sensitive ({\it HS}) detector output the signals from PMT anode come through a 30$^\times$ linear amplifier on an integrated circuit, while to the midi-sensitive one ({\it MS}) they are connected without any additional electronic amplification. At last, the pulses from an intermediate 6th PMT dinode which have a reduced amplitude and a reversed (positive) pulse polarity are transmitted to registration center just via the cable of the high-voltage PMT feeding through a capacitive AC-coupling to provide the third, low-sensitive ({\it LS}), informational channel. Such a three-diapason connection scheme of particle detectors permits to avoid any signal saturation which is quite common for the region of EAS core, and ensures the overall dynamic range of linear particle density measurements about $(3-7)\cdot 10^4$ (for combination of two {\it HS} and {\it MS} signals only), and up to $(1-2)\cdot 10^6$ with the use of all three amplitude diapasons.

In the room of data registration center the {\it HS} and {\it MS} diapason lines are connected to a multichannel inverting amplifier board where the polarity of scintillation signals is changed into a positive one, and further on they come to the board of analog memory circuits. The {\it LS} diapason signals are connected to the memory board through emitter follower type cascades without neither additional amplification nor phase inversion.

The system of analog memory is built on the SMP04 type sample-and-hold amplifier chips \cite{smp04}; it can keep the momentary amplitude of a pulse signal which has been present at its input in the arrival moment of a special control signal --- the trigger. Constant positive levels of scintillation amplitudes kept in the memory afterwards can be digitized with the multichannel system of amplitude-to-digital conversion (ADC) built on the basis of 8-channel, 12-bit AD7888 type integrated circuits \cite{ad7888}. The massive of digitized amplitude data accepted in every registered event is loaded into the common database table for further off-line analysis. At that, just three ADC codes, each varying in the range of 0-4095 are kept for every scintillation detector correspondingly to its three amplitude diapasons.

The multichannel ADC system is constituted by a set of separate modules with 128~informational conversion channels in each module. At the moment of a cosmic ray event registration all AD7888 chips in all ADC modules are operating simultaneously being synchronized by the local in-build controller. Due to such modular organization the time of event registration is quite negligible ($\le$0.5~ms), and does not depend on the total amount of conversion channels in the whole detector system, so the momentary pulse amplitude levels which have been latched at the outputs of analog memory unit can be considered as constant. After digitizing, the amplitude codes are temporarily kept inside the internal buffer memory of ADC module until they are requested by the side of central computer which controls the whole detector system. The subsequent data sending and storage takes the dead time about 0.1-0.3s after which the system is ready again for registration of the next event.

Besides the general multichannel ADC, the signals of the {\it HS} amplitude diapason are additionally connected to the pulse spectrometer and ratemeter subsystems. Twice a day, the analog commutator of the pulse spectrometer unit in turn switches any input signal to a single-channel pulse ADC for a fixed exposition time, so the amplitude spectra of the background scintillations can be regularly obtained for every shower detector. The position of the minimum ionizing particle peak (m.i.p.) in these spectra is used by individual fine tuning of the PMT high voltage feeding of scintillation channels, and later on it gives a base for the absolute detector calibration. Another set of analog pulse discriminator/digital counter pairs is used as a ratemeter for continuous registration of the current background count intensity in all scintillation channels (typically, with a 10s time resolution). These data are primarily intended for the real time control of detector operation stability.

Analogously, the {\it MS} diapason signals come to the subsystem of trigger generation. In this unit, the scintillation pulses from all detector points are summed together with analog circuits based on a set of CA3130E type operational amplifiers, and in the moment when the amplitude of the sum signal occurs above some predefined discriminator threshold, it is fired a shower trigger pulse. The latter is used for synchronization of data registration processes on all constituent parts of the EAS detector complex, and in particular, it causes latching of the momentary amplitude values in analog memory of amplitude channels. To provide the possibility of different logic of trigger generation the trigger subsystem permits both to sum the outputs of the whole set of scintillation detectors at once, and to sum them separately for a number of partial detector subsets, 12~scintillation channels per every group, keeping the information on the type of generated trigger in each case.

\section{The algorithm of shower data operation}
\label{sedaq}

The final goal of scintillation amplitude data operation applied to every EAS event registered at the Tien~Shan shower detector system is determination of basic EAS characteristics: the position of shower axis in installation plane, and the shower {\it age} $s$ and {\it size} $N_e$. The latter parameter has the physical meaning of the total number of shower particles, and it is nearly proportional to the energy of the primary cosmic ray interaction $E_0$. Besides, the zenith $\theta$ and azimuth $\phi$ directional angles of shower axis are defined on the basis of time delays of scintillation signal in a number of mutually spaced detectors.

\paragraph{The scintillation amplitude}

\begin{figure}
\begin{center}
\includegraphics[width=0.49\textwidth, clip,trim=0mm 0mm 0mm 0mm] {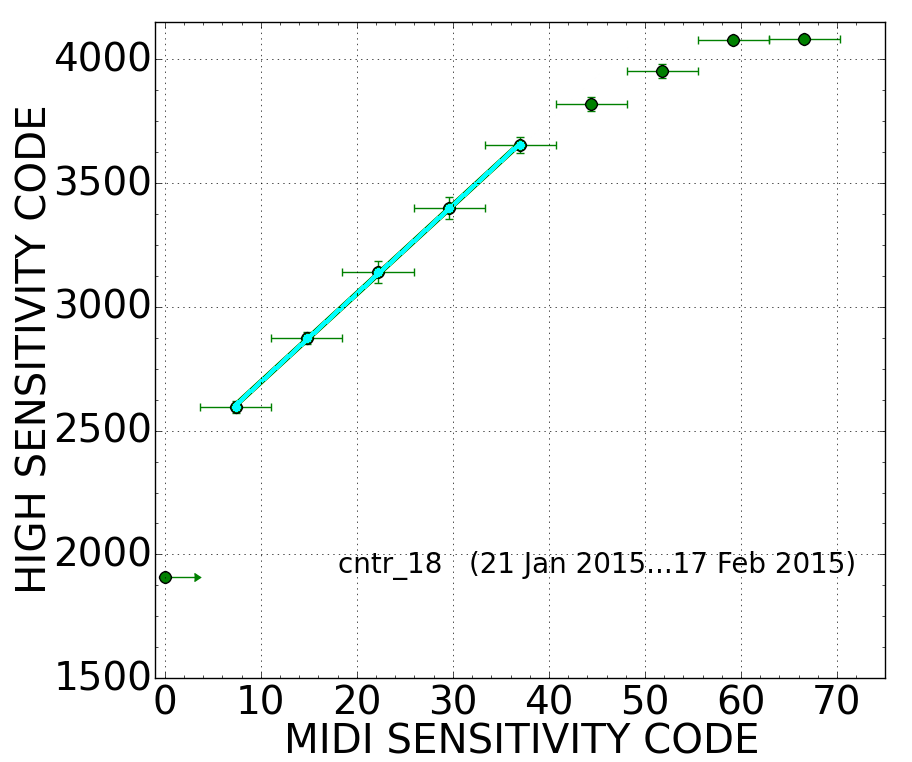}
\end{center}
\caption {Mutual "sewing" of the {\it HS} and {\it MS} diapason scales for a particular scintillation detector. Points --- the means of ADC codes registered in selected EAS events, line --- an approximation drawn over a saturation-free overlap range of the {\it HS} and {\it MS} scales.}
\label{figisewi}
\end{figure}

The first step of data operation consists in "sewing" together all partial measurements made initially in three ADC sensitivity diapasons into a single code of scintillation amplitude $P$, which procedure should be done individually for each detector channel in every registered EAS event. For the first pair of amplitude measurements, {\it HS} and {\it MS},  this sewing can be based on natural cosmic ray events when some non-zero code $P_{ms}$  appears already at the beginning of {\it MS} diapason scale, while the corresponding $P_{hs}$ code at the upper end of {\it HS} scale is not yet big enough to lose its linearity because of the electronics saturation effects; the cases of this type are met quite abundantly among all registered events. The essential partial overlapping of high-end {\it HS} and low-end {\it MS} amplitude scales was ensured by the proper tuning of conversion gains in the cascade of inverting amplifier (see the block diagram of figure~\ref{figidaq}), which has been done individually for every informational channel of the data acquisition system before the measurement starts.

The events of the mentioned type were selected from the general bank of experimental data for every detector channel of scintillation shower system over some time (typically, 10-15~days of uninterruptable operation of shower system in stable conditions), the means $\overline{P_{hs}}$ were calculated over the events with close $P_{ms}$ values, and the correlation plots $\overline{P_{hs}}(\overline{P_{ms}})$ were drawn, again individually for every considered detector and in considered interval of time. A typical sample of such correlation plot for a particular detector of the {\it Center} scintillation system over 3-weeks period is presented in figure~\ref{figisewi}. As it can be seen in this plot, before the saturation effects of {\it HS} diapason come into action, the $P_{hs}(P_{ms})$ dependence is a quite linear function: ${P_{hs}}(P_{ms})=a_{ms}\cdot P_{ms}+b_{ms}$, and both $a_{ms}$ and $b_{ms}$ parameters can be conveniently defined through approximation of experimental points by the least-squares method. The third parameter used in the sewing of amplitude diapasons, $t_{ms}$, is a threshold $P_{ms}$ value above which the {\it MS} diapason data start to be used for calculation of the unified amplitude code instead of original {\it HS} ones; usually, it is set at a maximum possible point above which the saturation of {\it HS} channel occurs quite visible. Hence, for the case presented in the discussed plot of the figure~\ref{figisewi} $a_{ms}=36$, $b_{ms}=2300$, and $t_{ms}=35$.

With accepted parametrization, the formal calculation rule of unified scintillation amplitude code $P$ over the first pair of sensitivity diapasons is $P=P_{hs}$~if~$P_{ms}<t_{ms}$, and $P=a_{ms}\cdot P_{ms}+b_{ms}$~if otherwise.

During the measurements, the sets of sewing parameters ($a_{ms}$, $b_{ms}$, $t_{ms}$) were periodically re-calculated in semi-automatic manner for every detector used, and stored in the special journal table of the common database.

The sewing of the next diapason pair, the {\it MS} and {\it LS} ones, is made quite analogously to the one described above. Because of the rarity of natural cosmic ray events with the non-zero {\it LS} amplitude code, the sewing of the third diapason is made on the basis of artificial data obtained when the interior of detector's reflector was illuminated with a short light flash emitted by a blue light-emitting diode. The electrical pulses generated to fire the LED were also initiating the procedure of data taken at ADC system in place of shower trigger, and the overlapping part of {\it MS} and {\it LS} scale codes obtained in these measurements permitted to define another pair of sewing coefficients $a_{ls}$ and $b_{ls}$.

\begin{figure}
\begin{center}
\includegraphics[width=0.49\textwidth, clip,trim=0mm 0mm 0mm 0mm] {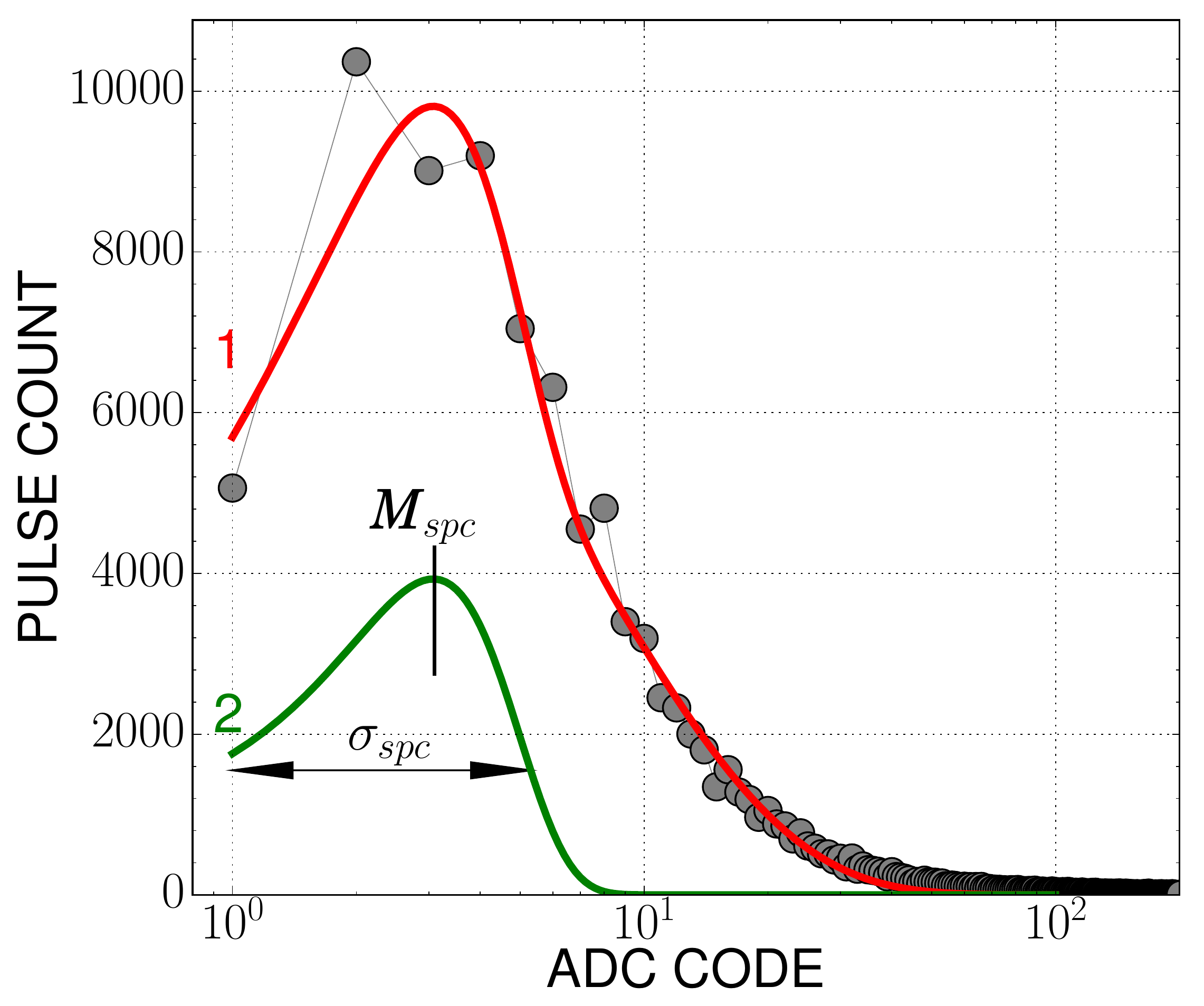}
\end{center}
\caption {Absolute calibration of scintillation detector based on the background m.i.p. amplitude spectrum (see text). Circles: experimental measurement points; curve {\it 1} --- their approximation with a sum of Gaussian and the power law function (see text); curve {\it 2} --- the pure Gaussian m.i.p. deposit.}
\label{figispecti}
\end{figure}

\paragraph{The density of EAS particles flow}
As the initial amplitude measurements $P_{hs}$, $P_{ms}$, and $P_{ls}$ occur being reduced to a single ADC scale, and the resulting unified amplitude code  $P$ is defined, it is possible to calculate the particle density distribution over the detectors of shower scintillation system. For this purpose, the spectra of background scintillation pulses are used which have been registered by turn regularly for every detector with the pulse spectrometer unit, as it is shown in figure~\ref{figidaq}. As a rule, these spectra have a characteristic shape with a single maximum in relative distribution of their intensity, like an example spectrum presented in figure~\ref{figispecti}. The position of this maximum $M_{spc}$ on abscissa scale can be interpreted as an average ADC code which corresponds to the passage of a fast single-charged particle through the scintillator, since relativistic electrons are known to be the most abundant component of the cosmic ray background (the minimum ionizing particles, m.i.p.). Before the measurements, the operation point of all detectors has been set by such a way to place the m.i.p. peak in their background count spectra to the beginning of the ADC code axis, usually somewhere in the interval between the codes 3-7. During the measurements, the current position of operation point is constantly checked by the measurement of the background pulse spectra and regulated through tuning of the PMT high voltage if necessary.

Since the overwhelming part of charged particles constituting an EAS does also consist of relativistic electrons, the average number of particles which have gone through a detector at the moment of shower event can be defined as  $P/M_{spc}$, and its relation to the sensitive area of scintillator $S$ gives an estimation of the density of particle flow: $\rho=P/(M_{spc}\cdot S)$.

Routinely, the normalization factor $M_{spc}$ is defined and checked for every scintillation detector by the analysis of its background pulse spectra. To make this procedure semi-automatic, each experimental spectrum of ADC codes $x$ is approximated with a sum $G( x, M_{spc}, \sigma_{spc} ) + a / x^b $ of a Gaussian $G$ which corresponds to the regular m.i.p. deposit, and a power law function responsible to describe stochastic noise spectrum of electronic channel; the Gaussian parameters $M_{spc}$ and $\sigma_{spc}$, so as the coefficients $a$ and $b$ being defined properly through the fit of experimental points. In the figure~\ref{figispecti} the fitting function of such a kind is represented by the curve{\it~1}, while the curve{\it~2} indicates the pure Gaussian deposit only into this approximation. The value of $M_{spc}$ parameter which would be accepted in the case of this spectrum is~3.0, and $\sigma_{spc}=2.3$.

Usually at the time of EAS registration, parameters $M_{spc}$ and $\sigma_{spc}$ are checked and re-defined daily in the way discussed here for all detectors of shower system; their estimations are kept in a special table of general database and selected automatically by calculation of the particle density distributions $D(\rho)$ in all registered shower events.

\paragraph{Dynamic range of the particle density measurements}

\begin{figure}
\begin{center}
\includegraphics[width=0.49\textwidth, clip,trim=0mm 0mm 0mm 0mm] {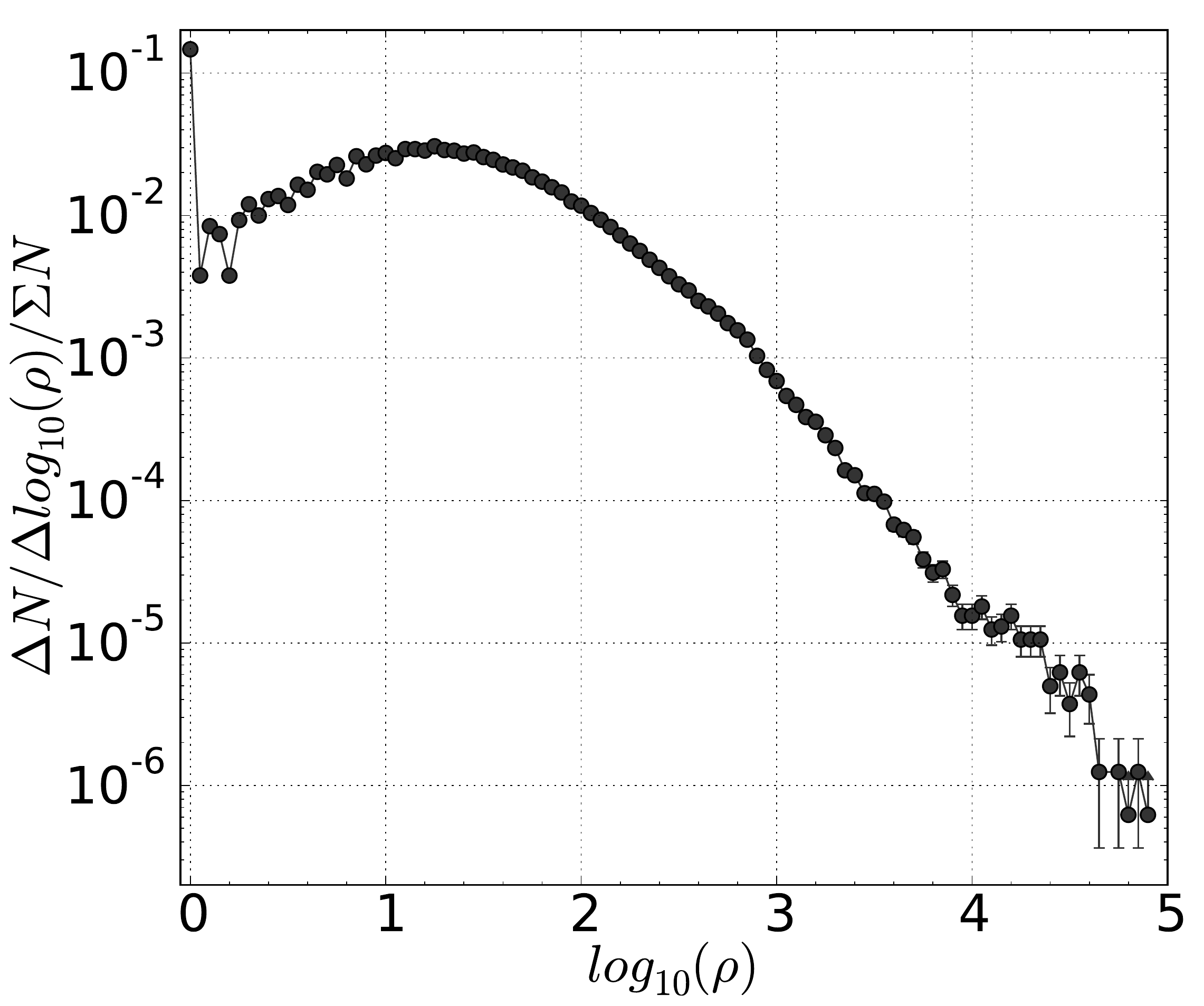}
\end{center}
\caption {Typical distribution of experimental particle density $\rho$ estimations summed over a one-week long time period.}
\label{figidensi}
\end{figure}

The massive of particle density estimations $\rho$ obtained in every registered EAS event according to the procedure described above can be used for regular check of the real dynamic range of density measurements which is practically accessible for shower system. For this purpose all events are selected from the database over a rather prolonged period of time (usually, a week), and the distribution of particle density estimations $\rho$ registered in considered events set is calculated for all detectors. The general shape of this distribution, particularly absence of any irregularities at the high end of the $\rho$ axis, is a sign of normal operation of the considered particle detector, and also the verification of the correctness of particle density calculation procedure accepted in the current experiment as a whole.

A sample of particle density plots of the described type is presented in the figure~\ref{figidensi}, where all partial distributions accepted for every particular detector are additionally summed together to obtain a possibly noticeable statistics in the high range of $\rho$ values. In calculation of these distributions only the data of two leading amplitude diapasons {\it HS} and {\it MS} were used. It is seen that in the range of high density values $\rho\ge 10^{4}$ the experimental distribution diminishes regularly, without any sharp cut-offs; this is an evidence that even the use of two amplitude diapasons is sufficient to gain a dynamic range of particle density measurements of the order of $(5-7)\cdot 10^4$.  According to the properties of lateral distribution function of particle density in a typical EAS (see the next paragraph), this is quite enough for immediate investigation of the EAS core region up to the showers with total charged particles multiplicity (shower {\it size}) $N_e\sim (1-2)\cdot 10^7$, or, using an estimation commonly accepted for the Tien~Shan station's height, $E_0/1$eV$\sim (2-3)\cdot 10^9 N_e$ \cite{oneasbook}, up to primary EAS energy $E_0$ about $(3-5)\cdot 10^{16}$~eV.

Evidently, an additional factor of $20-30$ which remains to achieve a $10^6$~range of density measurement which is necessary for the study of higher EAS energies, up to $E_0\sim 10^{18}$~eV, can be obtained with applying the third amplitude diapason, although it has not still been verified in practical measurements. Nevertheless, all the electronics wanted for this purpose according to the block diagram of figure~\ref{figidaq} has been just mounted in its operation place and is now in a ready-to-use state. The task of {\it LS} diapason measurements is planned to be done in nearest future, together with commissioning of the peripheral part of scintillation shower system, which is needed, again, to increase the size of shower installation up to the area necessary for reliable registration and analysis of the high-energy EAS.

\paragraph{Determination of the EAS parameters}

\begin{figure*}
\begin{center}
\includegraphics[width=0.49\textwidth, clip,trim=0mm 0mm 0mm 0mm] {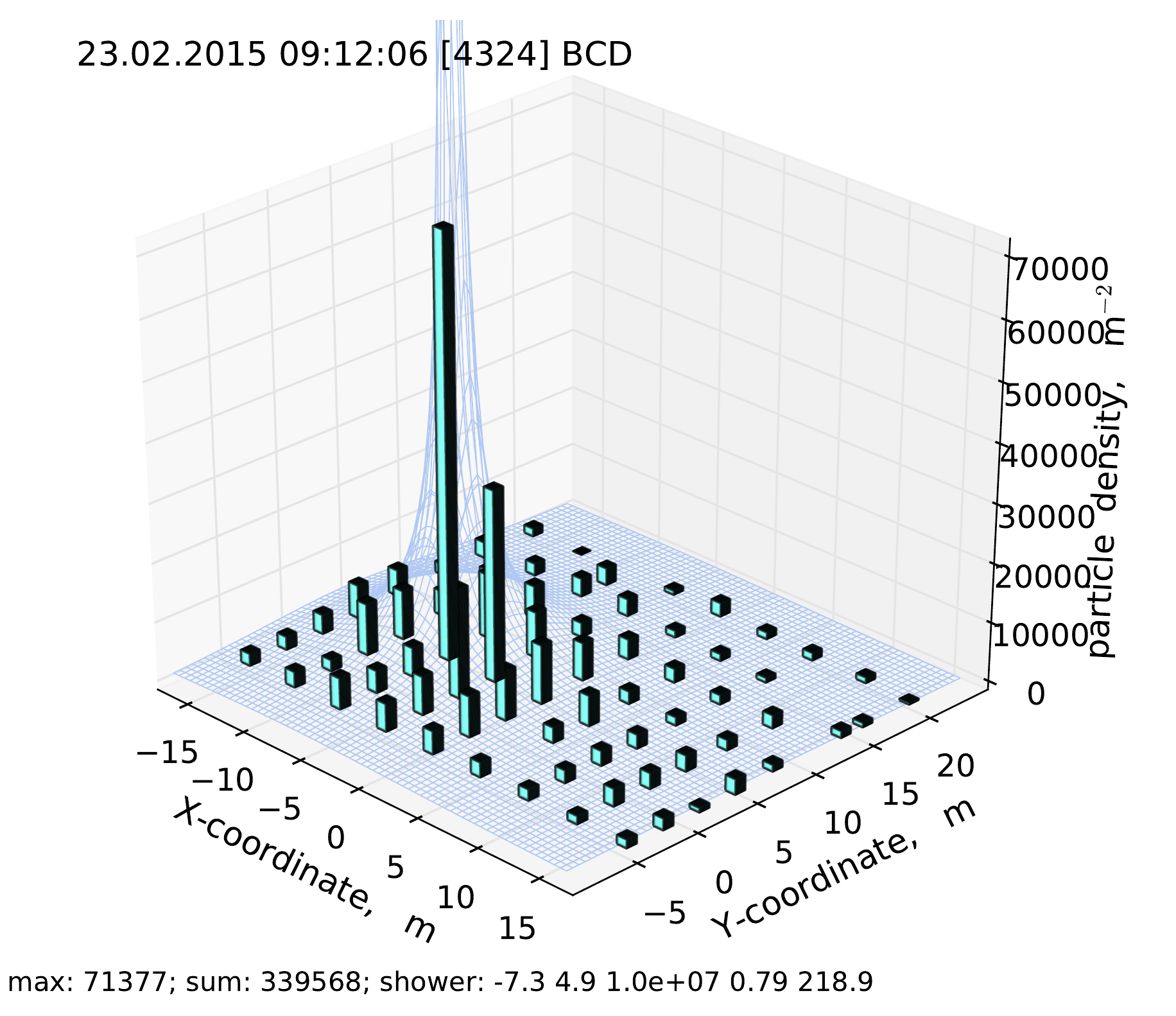}
\includegraphics[width=0.49\textwidth, clip,trim=0mm 0mm 0mm 0mm] {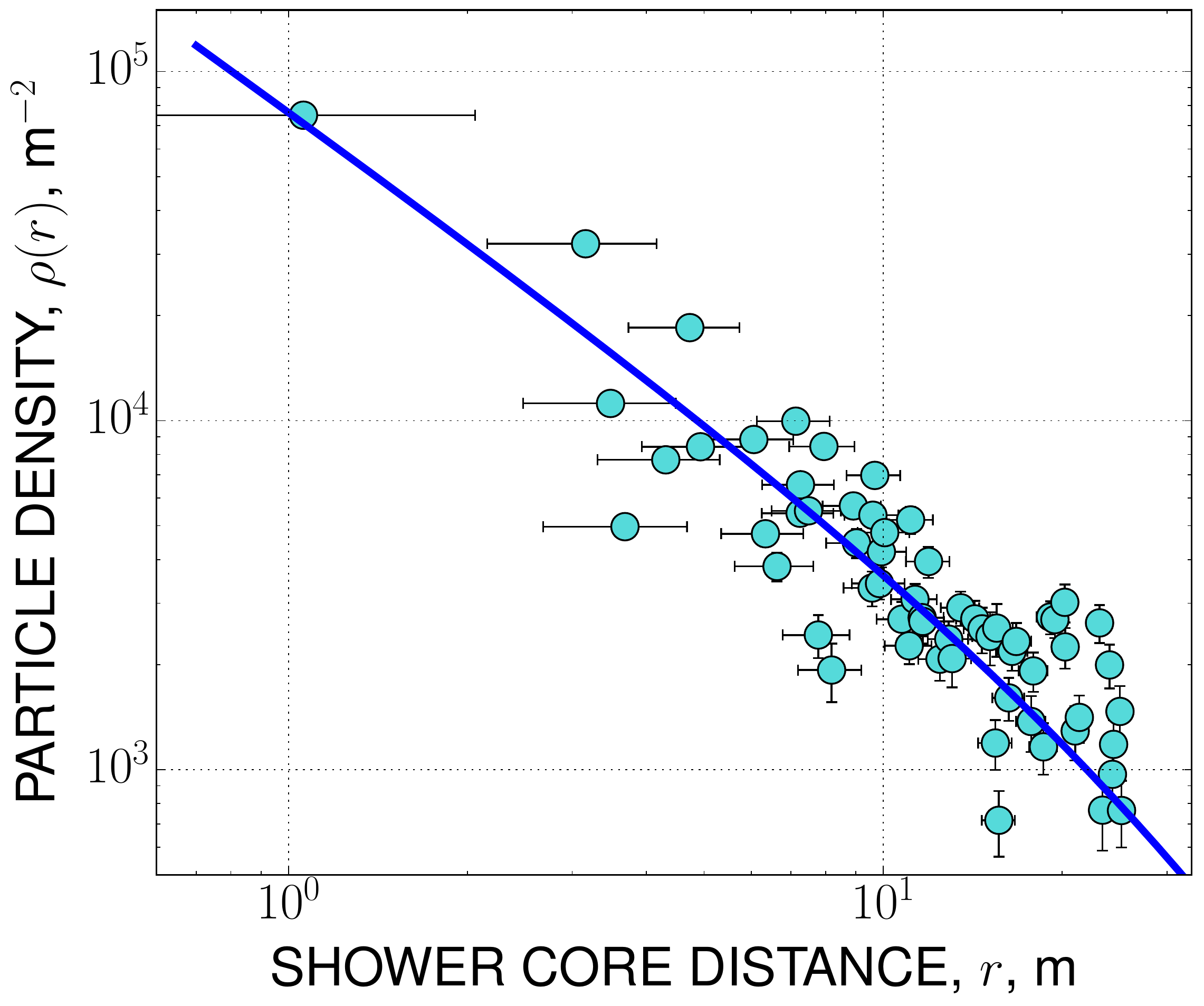}
\end{center}
\caption {Left plot: approximation of the experimental density distribution with a two-dimensional NKG function in a typical EAS event. Right plot: the lateral distribution of particle density measured by scintillation detectors in the same event (circles), and its NKG approximation (line).}
\label{figiappri}
\end{figure*}

As the spatial distribution of particle density $D(\rho)$ is known for a shower event, it is possible to define basic characteristics of the registered EAS: the shower {\it size} $N_e$, its {\it age} parameter $s$, and the location of shower axis $(x_0,y_0)$ in the coordinate frame of detector system. These parameters are defined through minimization of a $\chi^2$ type functional
\[
\begin{array}{c}
\chi^2=\sum_D{ (\frac{\rho_D - \rho_{NKG}( r_D( x_0, y_0 ), s, N_e ) }{\sigma( \rho_D ) } )^2 }\\
\rightarrow \min_{[x_0, y_0, s, N_e]},
\end{array}
\]
where $\rho_D$ is the particle density registered in detector point $D$;  $\rho_{NKG}$ --- the value of theoretical density distribution 
at the distance $r_D$ between the location of considered detector $D$ and the center point $(x_0,y_0)$ of an EAS with shower parameters $(s, N_e)$; and the summing is made over the whole set of detectors used. The differences between the experimental and theoretical density values are supposed to be normalized to the error of particle density estimation $\sigma(\rho_D)$ which is calculated with the use of ($M_{spc}$, $\sigma_{spc}$) pair of parameters defined in Gaussian approximation of the background m.i.p. amplitude spectrum of considered detector $D$ (see the figure~\ref{figispecti}).

Theoretical distribution of the shower particles density is parametrized in the form of Nishimura--Kamata--Greisen function \cite{nkg-greisen,nkg-nk}:
\[
\begin{array}{c}
\rho_{NKG}( r, s, N_e )=
0.366 s^2 ( 2.07-s )^{1.25}\times \\
( r / R_M )^{s - 2 }( 1 + r / R_M )^{s - 4.5} / R_M^2 )\times N_e,
\end{array}
\]
with the value of Mollier radius $R_M=120$m at the altitude of Tien~Shan station.

Minimization of the $\chi^2$ functional is made with the use of the standard  procedure taken from optimization module of the SciPy mathematical library \cite{scipy} which implements the Nelder-Mead \cite{neldermead} algorithm for the search of global extremum on a function profile. A typical result of such approximation of the experimental 2D particle density distribution in an EAS event is illustrated by the figure~\ref{figiappri}.

\paragraph{Physical results of the test measurements}

\begin{figure*}
\begin{minipage}{0.45\textwidth}
\includegraphics[width=1\textwidth, trim=0mm 0mm 0mm 20mm]{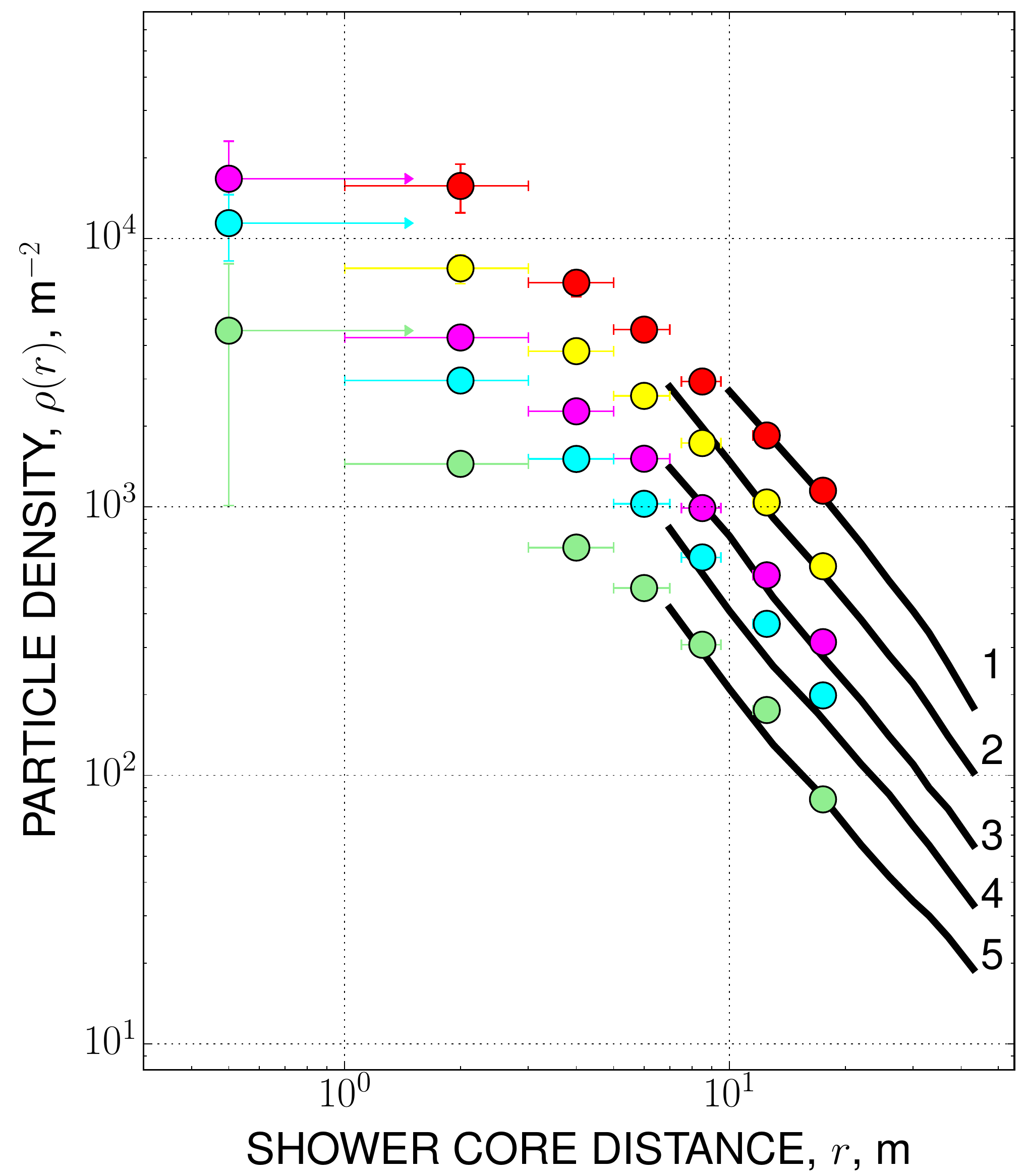}
\end{minipage}
\begin{minipage}{0.55\textwidth}
\includegraphics[width=1\textwidth, trim=0mm 0mm 0mm 20mm]{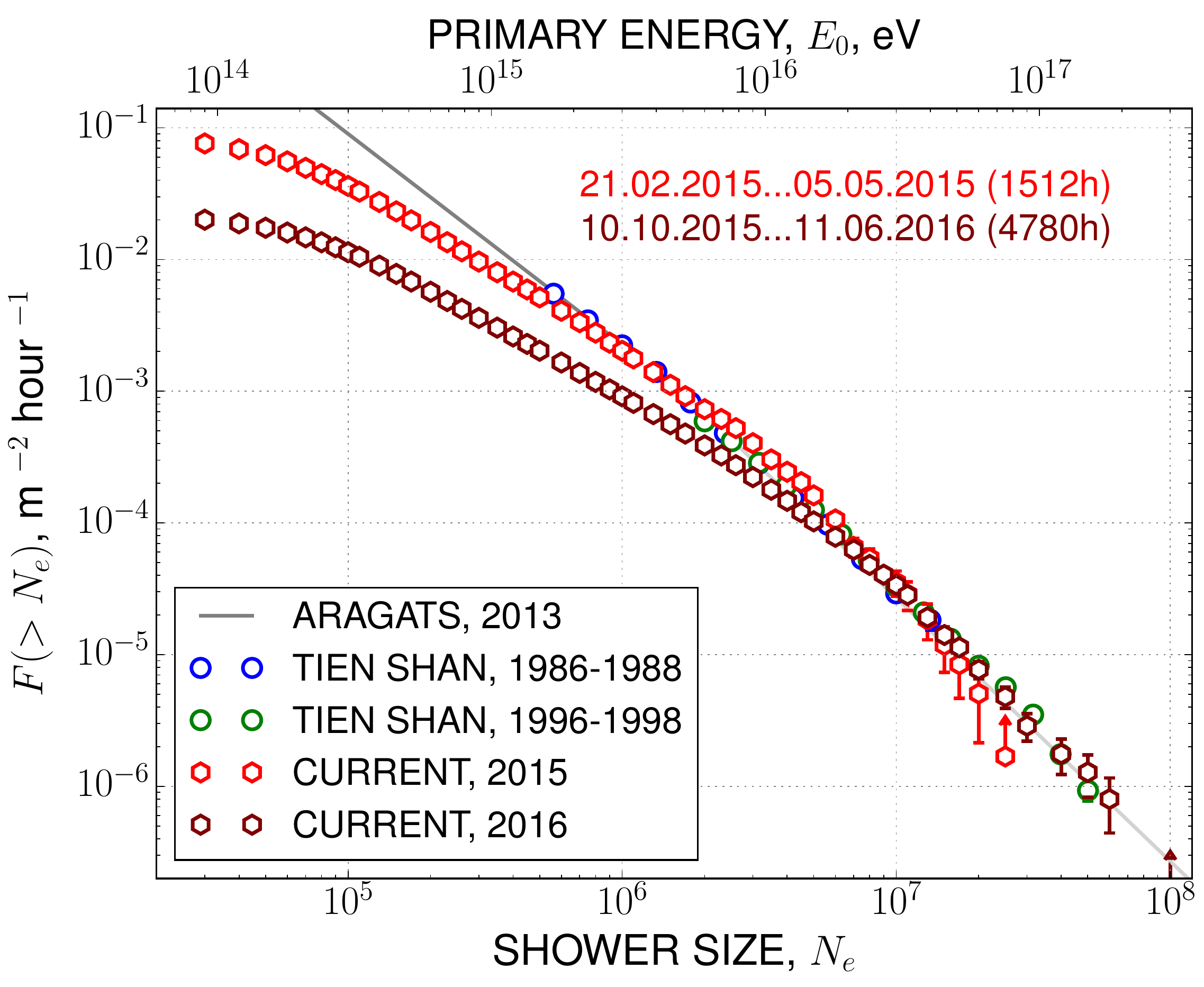}
\end{minipage}
\caption{
Left: the lateral distribution functions of particle density in EAS with $N_e=7.3\cdot 10^6$ ({\it 1}), $4.1\cdot 10^6$ ({\it 2}), $2.3\cdot 10^6$ ({\it 3}), $1.3\cdot 10^6$  ({\it 4}), and $7.3\cdot 10^5$({\it 5}); both measured in the current (circles) and Aragats experiment \cite{aragats_fpr_b} (lines). Right: the integral EAS size spectrum obtained at the new shower installation (red circles) in comparison with the spectra measured earlier at Tien~Shan \cite{jopg2001,hadron_spc} (green and blue circles), and Aragats \cite{aragats_spc} (gray line). For convenience, the upper axis presents the estimations of corresponding primary energy $E_0$ which have been calculated as $E_0=3\cdot 10^9 N_e$.}
\label{figiscintiphysi}
\end{figure*}

The first stage of the shower system, i.e. the {\it Center} scintillation carpet, was operating continuously, and the data taking procedure described above was held in its full volume since December~2014 until May~2015. The results collected in this period of test exploitation and their comparison with the known data on EAS characteristics are discussed here to verify the correctness of hardware functioning and adequacy of the data operation algorithms. It should be noted that neither additional peripheral detectors nor any angular information on the direction of primary cosmic ray particles were available so far, and only two diapasons of the shower particle density measurements, i.e. {\it HS} and {\it MS}, were used in this test run.

The left plot on the figure~\ref{figiscintiphysi} presents the lateral distribution functions of the density of EAS charged particles, with coloured circles denoting the data currently obtained in the present work, and with black continuous lines --- those published for corresponding intervals of the shower size by GAMMA experiment \cite{aragats_fpr_b} (the latter one being situated at Mount Aragats at the same altitude as the Tien~Shan station and using similar detectors and electronic equipment for EAS registration grants a convenient reference for valuation of the reliability of our experimental set-up). In the range of shower core distances where the distributions of both experiments do overlap it is seen a rather close agreement of all data sets, both in their absolute intensity, and in distribution slope. Absence of any rough deviations of the registered distributions in the range of small core distances $r\lesssim 1$m is another verification of the fact that the accepted way of particle density restoration does indeed ensure its announced dynamical range of the order of $(3-7)\cdot 10^4$, even with the use of two amplitude diapasons only.

The integral spectra of the registered EAS over the number of charged particles $N_e$ are shown in the right plot of figure~\ref{figiscintiphysi}, where again the spectrum obtained in current experiment (red points) is compared with the ones measured at different times by the old shower detector system of the Tien~Shan mountain station (blue and green) \cite{jopg2001,hadron_spc}. The approximation of the Aragats spectrum published in \cite{aragats_spc} is shown on this plot also by the gray line. The spectrum newly obtained agrees satisfactorily with other ones, both in its absolute intensity and slope; its slight deviation from the correct pure power form in the range of $N_e\gtrsim 2\cdot 10^6$ can be explained by the absence of peripheral shower detectors in our test measurements which resulted in noticeable side effect with relative rise of reconstructed $N_e$ values when the lateral size of shower occurs comparable with geometrical size of the whole {\it Center} carpet. The deviation of the shower size spectrum in figure~\ref{figiscintiphysi} from the pure power law at its low-$N_e$ side indicates that the threshold of 100\% EAS registration probability at the time of considered test measurements was about $N_e\sim 2\cdot 10^5$ ($E_0\sim 5\cdot 10^{14}$eV) which can easily be changed later through corresponding tuning of the trigger generation system in future experiments, if necessary. 

Generally, the satisfactory agreement between the known and the newly obtained shower characteristics is the evidence of the correctness of the proposed scintillation data operation procedure.

\section{Neutron detectors of the nuclear-active cosmic ray component}
\label{seneu}

\paragraph{The neutron detector system of the Tien~Shan station}

\begin{figure*}
\begin{center}
\includegraphics[height=0.7\textwidth, angle=90, clip, trim=0mm 0mm 90mm 45mm]{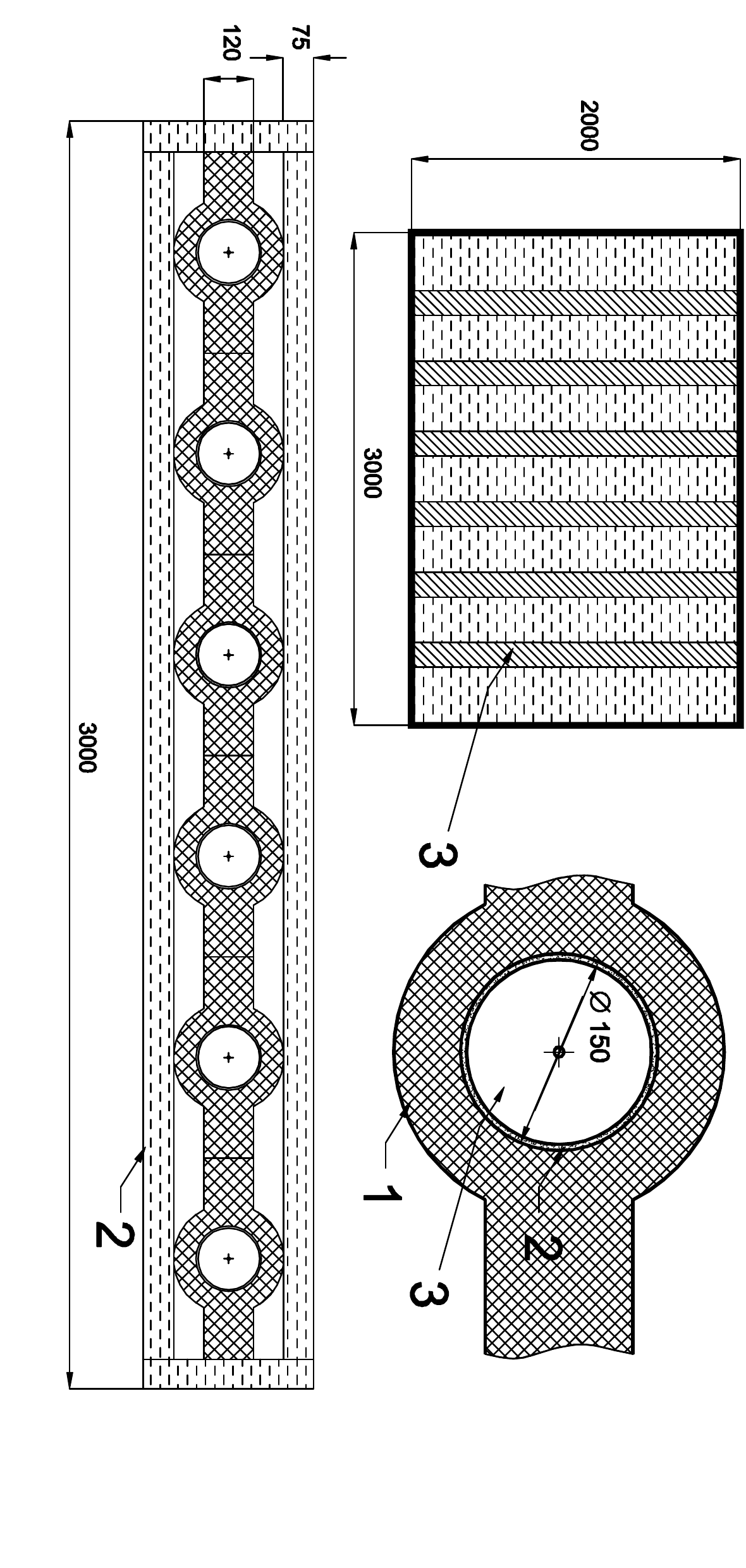}
\caption{Internal arrangement of the NM64 type neutron supermonitor unit ({\it 1} -- lead, {\it 2} -- polyethylene, {\it 3} -- the SNM15 type proportional neutron counter); dimensions are shown in mm.}
\label{figineumo}
\end{center}
\end{figure*}

Historically, the system of neutron registration at Tien~Shan has been developing around the 18NM64 type neutron supermonitor which was installed here many years ago, and for several decades has been participating in the world wide net of the cosmic ray intensity monitoring \cite{ontienmonitor}.

Tien~Shan supermonitor consists of three 2$\times$3m$^2$ standard units  \cite{carmichel_supermonitor} situated compactly in one building. As it is shown in schematic picture of figure~\ref{figineumo}, every unit contains six big proportional neutron counters, 2m long and 15cm in diameter. The counters are filled with a mixture of Ar and BF$_3$ gases, under partial pressure of 940~and 300mm~Hg correspondingly. The BF$_3$ gas filling of neutron counters is up to 90\% enriched with $^{10}$B isotope so the registration of {\it thermal} neutrons is possible due to nuclear reaction $n+^{10}$B$\rightarrow ^7$Li$+ \alpha$.

The neutron counters inside the supermonitor unit are surrounded with interleaving layers of 10cm thick lead absorber, and of light hydrogen-rich neutron moderator substance (the polyethylene moderator tubes with a 1cm wall thickness around each counter separately, and the 75cm thick common polyethylene shielding which covers the unit as a whole from outside). Due to such internal arrangement, the supermonitor is sensitive mainly to the flux of high-energy cosmic ray hadrons (above some hundreds of MeV) which are capable to produce a multitude of evaporation neutrons in interaction with nuclei of its heavy absorber. Later on, these secondary neutrons quickly (for the time of $\sim$10$\mu$s) loose their energy up to the thermal level ($\sim$10$^{-2}$eV) in multiple coincidences with light moderator nuclei, and diffuse inside the monitor until their escape to outside, a capture by hydrogen in moderator layers, or registration within one of neutron counters. The characteristic lifetime of evaporation neutrons is an order of magnitude higher than thermalization time, and for the NM64 supermonitor set-up it is about $600-650$$\mu$s. Besides the high-energy CR hadrons, the supermonitor unit can also register low-energy neutrons ($\sim$1MeV and below), though with a very small $0.05-1$\% probability. The outer polyethylene reflective shielding of the supermonitor prevents any excessive influence of external low-energy neutron background on the signal from CR interactions, and partly increases the net registration efficiency thanks to reflection of diffusing evaporation neutrons from within back in direction of the monitor's interior.

The average thickness of lead absorber within NM64 supermonitor unit is around 0.5 interaction length of energetic hadrons, so the registration probability of any particular CR hadron is about one half.

Neutron detectors as a mean for estimation of cosmic ray hadrons' energy were introduced at the beginning of 1950th \cite{simpson}, and their basic functional principles are the following. The number of evaporation neutrons $\nu$ born by hadron interaction with a heavy absorber nucleus inside the supermonitor unit depends on the energy of projectile hadron $E_h$. Of these neutrons, some have a chance to be registered in falling into one of neutron counters, so the number of registered neutron signals $M$ ({\it multiplicity}) is also a function of $E_h$. The means of both values are connected with obvious dependence $\langle M \rangle(E_h)=\epsilon \cdot \langle \nu \rangle(E_h)$, or inversely $E_h=E_h(M / \epsilon)$, where $\epsilon$ is the overall efficiency of supermonitor unit (i.e. the probability of neutron registration), and the averaging sign brackets are omitted in last expression.

The efficiency of neutron registration by the Tien~Shan 18NM64 neutron supermonitor was studied experimentally both {\it in situ} with the use of a calibrated Pu--Be neutron source, and with a special model set-up irradiated in hadron beams of Serpukhov accelerator \cite{inca1998c,inca1999d}. The function of energy response $E_h=E_h(M)$ in the range of high energies ($E_h\gtrsim$1GeV) was derived through comparison of the registered multiplicity spectrum of neutron events with the energy spectrum of cosmic ray hadrons which was known from preceding direct measurements at Tien~Shan calorimeter \cite{nmn2003}. Further on, this function was calculated in a much wider energy range through complete simulation of the neutron production and propagation processes inside a supermonitor unit made with the use of Geant4 simulation toolkit \cite{geant_base}.

By simulation, the Geant4 models of elastic coincidence of thermal, intermediate (4eV--20MeV), and high-energy (above 20MeV) neutrons were taken into account, together with the processes of radiative neutron capture, and with models of inelastic hadronic interaction of neutrons, protons, and pions in corresponding energy ranges. For protons and charged pions the processes of their multiple scattering and ionization losses were considered, and for pions the corresponding decay models; the physics of negative pions included the process of their absorption at rest. The set of electromagnetic processes included bremsstrahlung, multiple scattering, ionization losses of electrons and positrons (for positrons, the annihilation process as well). For gamma-radiation, there were used the models of photoelectric effect, Compton scattering, pair production, and photonuclear interactions.

\begin{figure}
\begin{center}
\includegraphics[width=0.5\textwidth, trim=0mm 0mm 0mm 0mm]{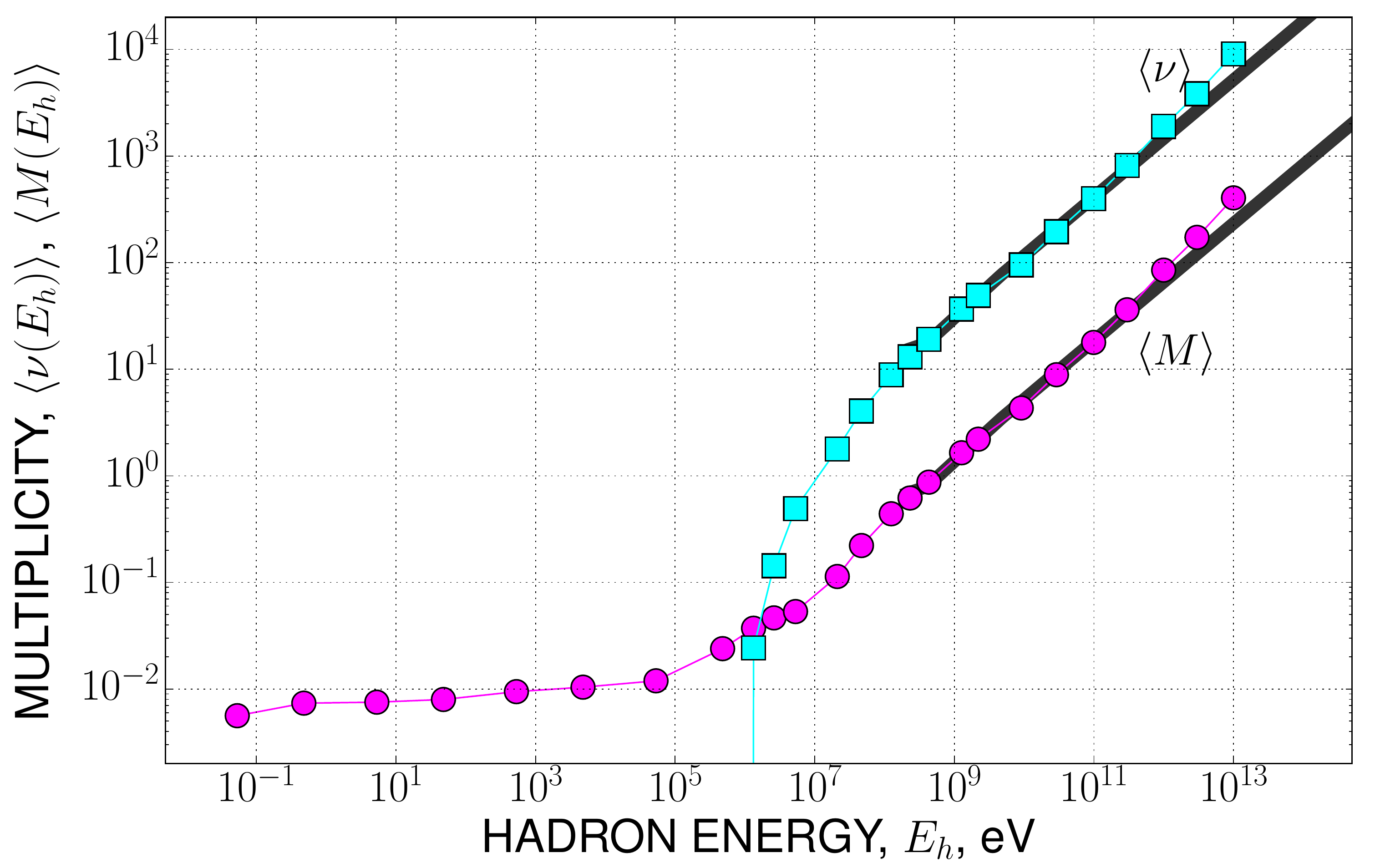}
\caption{The data of a Geant4 simulation of the energy dependence of mean number of evaporation neutrons $\langle \nu(E_h) \rangle$ and registered multiplicity of neutron counter signals $\langle M(E_h) \rangle$ for a standard NM64 unit (points) in comparison with the results of experimental measurements at Tien~Shan monitor (straight continuous lines).}
\label{figineumoeffi}
\end{center}
\end{figure}

The result of this work is presented in the figure~\ref{figineumoeffi}. It can be seen, that the use of NM64 type supermonitor for detection of cosmic ray hadrons starts to be effective ($\langle M \rangle \gtrsim 1$) at the energy $E_h \sim 0.5-1$GeV; later on, its energy dependence on registered signal is nearly quadratic: $E_h=0.49\cdot M^{1.8}$GeV~\cite{nmn2003} which formula is valid in a wide range of hadron energies $E_h=1-10^5$GeV. This conclusion does correspond to the previous results of early works \cite{hughes_supermonitor,onmulti}.

It is also seen a good agreement in the plot of figure~\ref{figineumoeffi} between the data of Geant4 simulation and experimental CR measurements in the overlapping range of high $E_h$ values. This is the evidence of correctness of the physical models accepted in simulation, and of general reliability of its result.

\begin{figure*}
\begin{center}
\includegraphics[width=0.49\textwidth, trim=20mm 0mm 20mm 70mm]{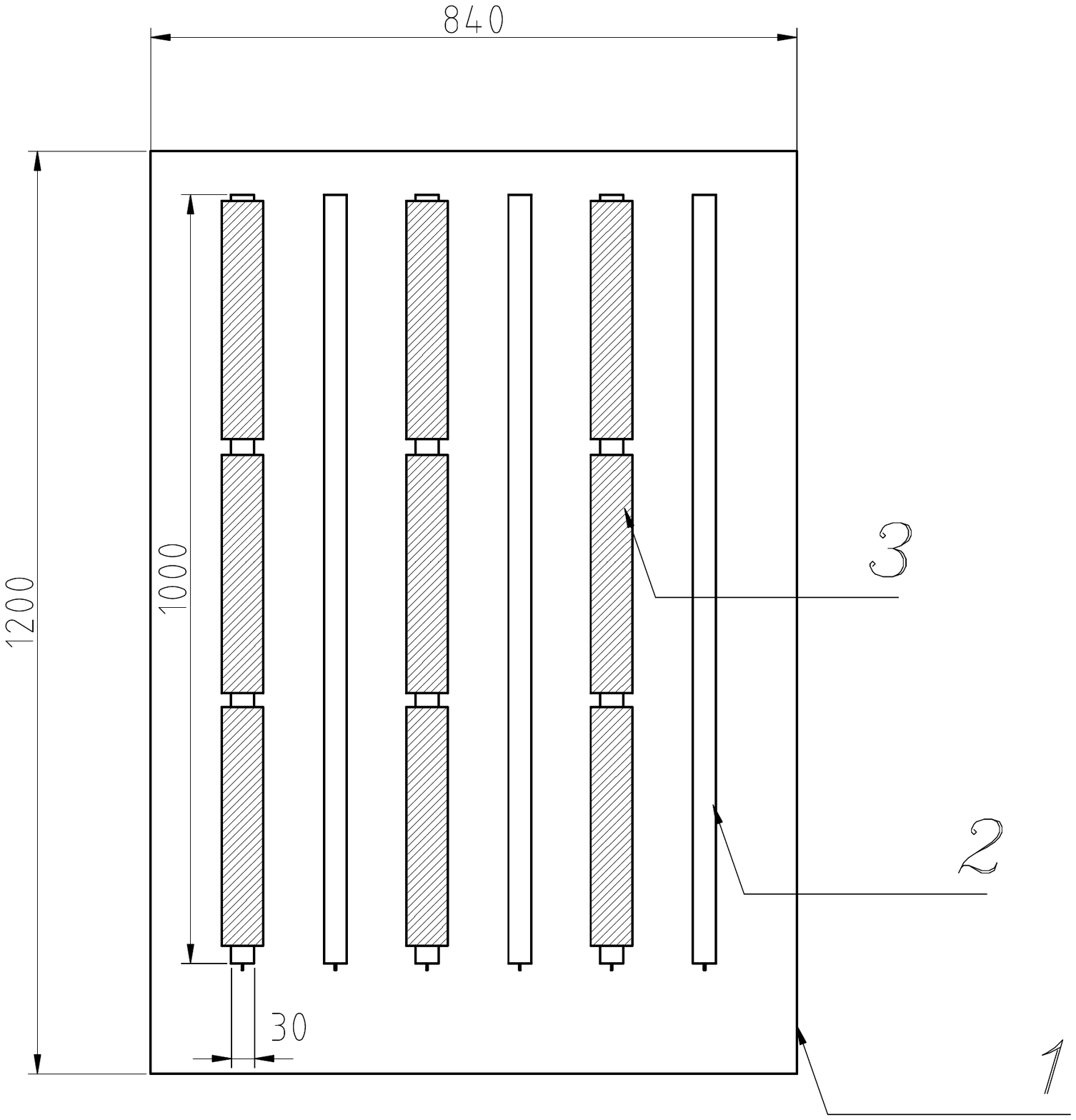}
\includegraphics[width=0.49\textwidth, trim=0mm 0mm 0mm 0mm]{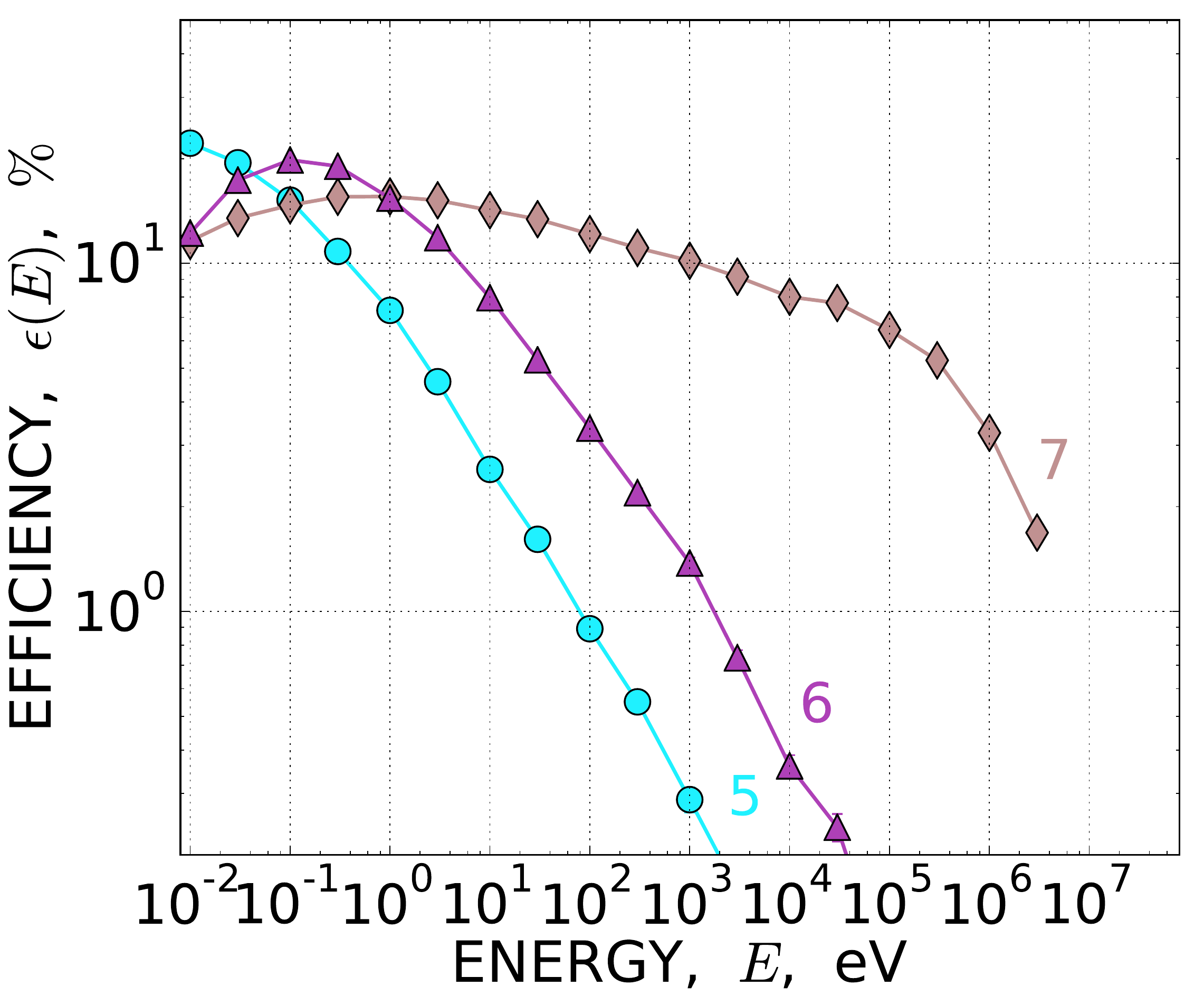}
\caption{Left: the low- and intermediate energy neutron detector ({\it 1} -- 1mm thick wall of an outer aluminum box,  {\it 2} -- the proportional neutron counter, {\it 3} -- polyethylene moderator tubes). Right: the energy dependence of registration efficiency with different types of neutron detectors (Geant4 simulation; {\it 1} -- for a 6 neutron counters box without any moderator as a whole,  {\it 2} and {\it 3} -- for a single counter inside a paraffin tube with the wall thickness of 0.5cm and 1cm; {\it 4} -- for the NM64 neutron monitor. By calculation of the low-energy neutron detector efficiencies the surrounding environment was taken into account).}
\label{figineudete}
\end{center}
\end{figure*}

To register low-energy neutron accompaniment of EAS passage at the Tien~Shan station it is used a special type of detector set-up \cite{jopg2008}. A detector of this kind consists of an aluminum box with 6~proportional neutron counters placed inside (see the figure~\ref{figineudete}). The counters are filled with $^3$He gas under the pressure of 2~atmospheres, so that the registration of thermal neutrons is possible due to reaction $n+^3$He$\rightarrow ^3$H$+ p$. Plain counters without any ambient moderator coating, and the counters covered with moderator tubes of different thickness are both applied to gain sensitivity in various ranges of neutron energy. The energy dependence of their efficiency which is shown on the plot of figure~\ref{figineudete} was defined by Geant4 simulation of the processes of neutron registration under different geometry conditions \cite{yanke2011}. The model of particle interaction used by this simulation was the same as by calculations of the NM64 supermonitor efficiency, but the accepted detector geometry module took into account also the influence of surrounding environment on resulting multiplicity of the registered neutron signals: the scattering of thermal neutrons in the air and in construct elements of the building around the detector, as well as in the soil beneath it.

\paragraph{Procedure of neutron data registration}

\begin{figure}
\begin{center}
\includegraphics[width=0.5\textwidth, trim=10mm 70mm 10mm 70mm]{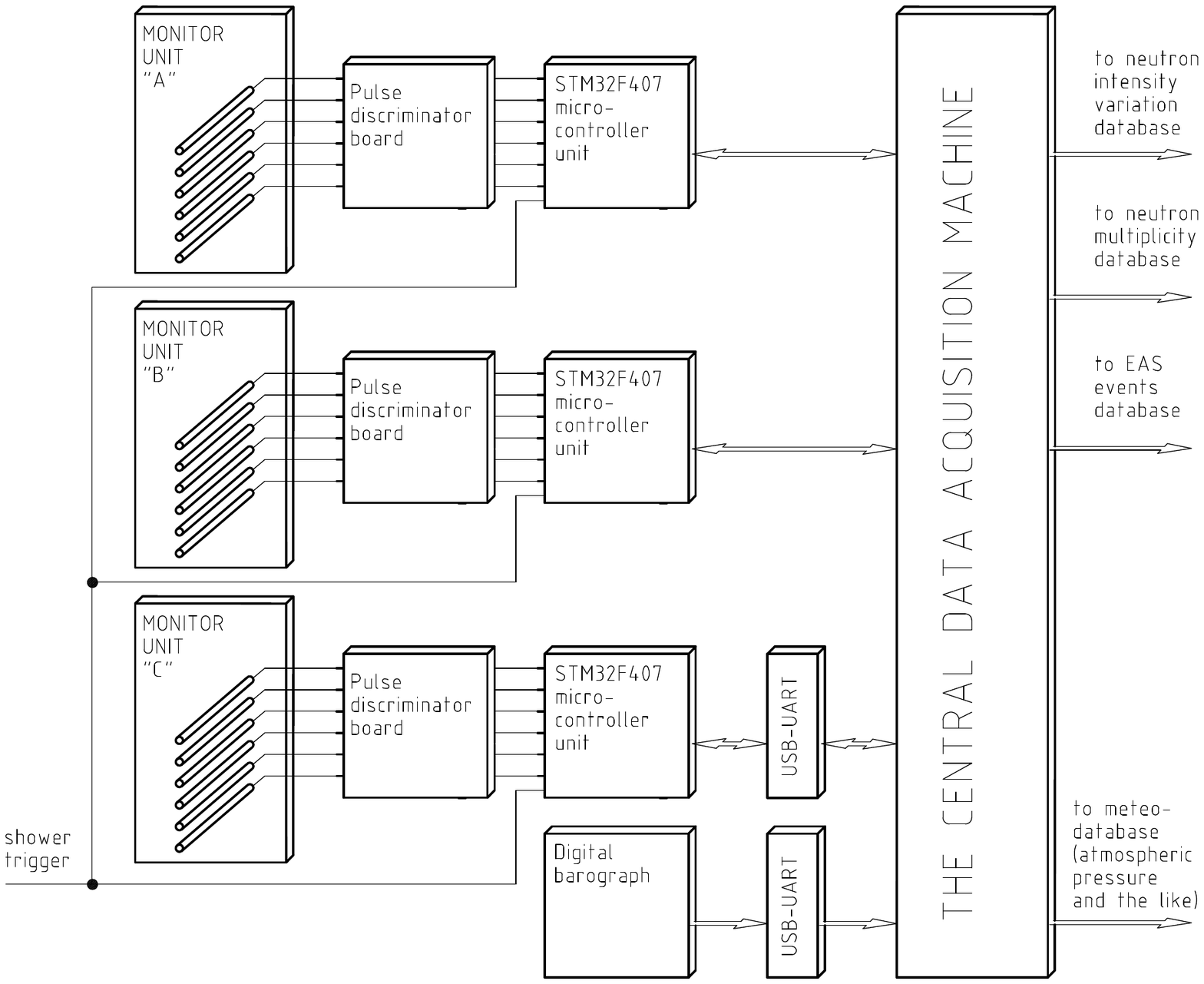}
\caption{Block diagram of the neutron data operation.}
\label{figineufuncti}
\end{center}
\end{figure}

\begin{figure}
\begin{center}
\mbox{
\includegraphics[width=0.50\textwidth, trim=0mm 0mm 0mm 0mm]{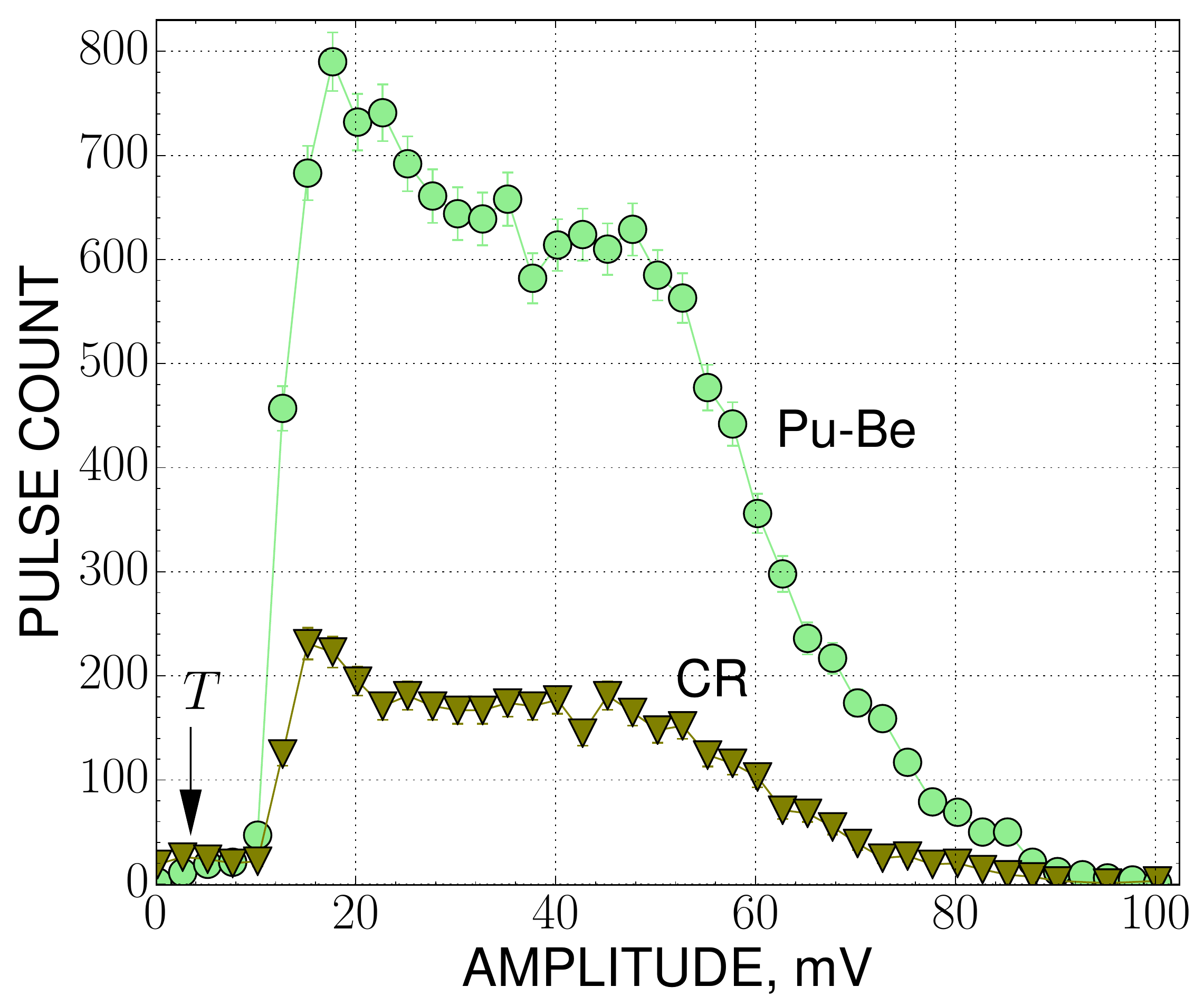}
}
\caption{The pulse amplitude spectra of a $^{10}$BF$_3$ filled neutron counter placed inside the NM64 supermonitor by registration of the pure CR hadrons, and in the presence of a Pu-Be neutron source. The {\it T} arrow indicates the position of the pulse discrimination threshold accepted in the channel of data acquisition.}
\label{figineuspecti}
\end{center}
\end{figure}

\begin{figure}
\begin{center}
\mbox{
\includegraphics[width=0.48\textwidth, trim=0mm 0mm 0mm 0mm]{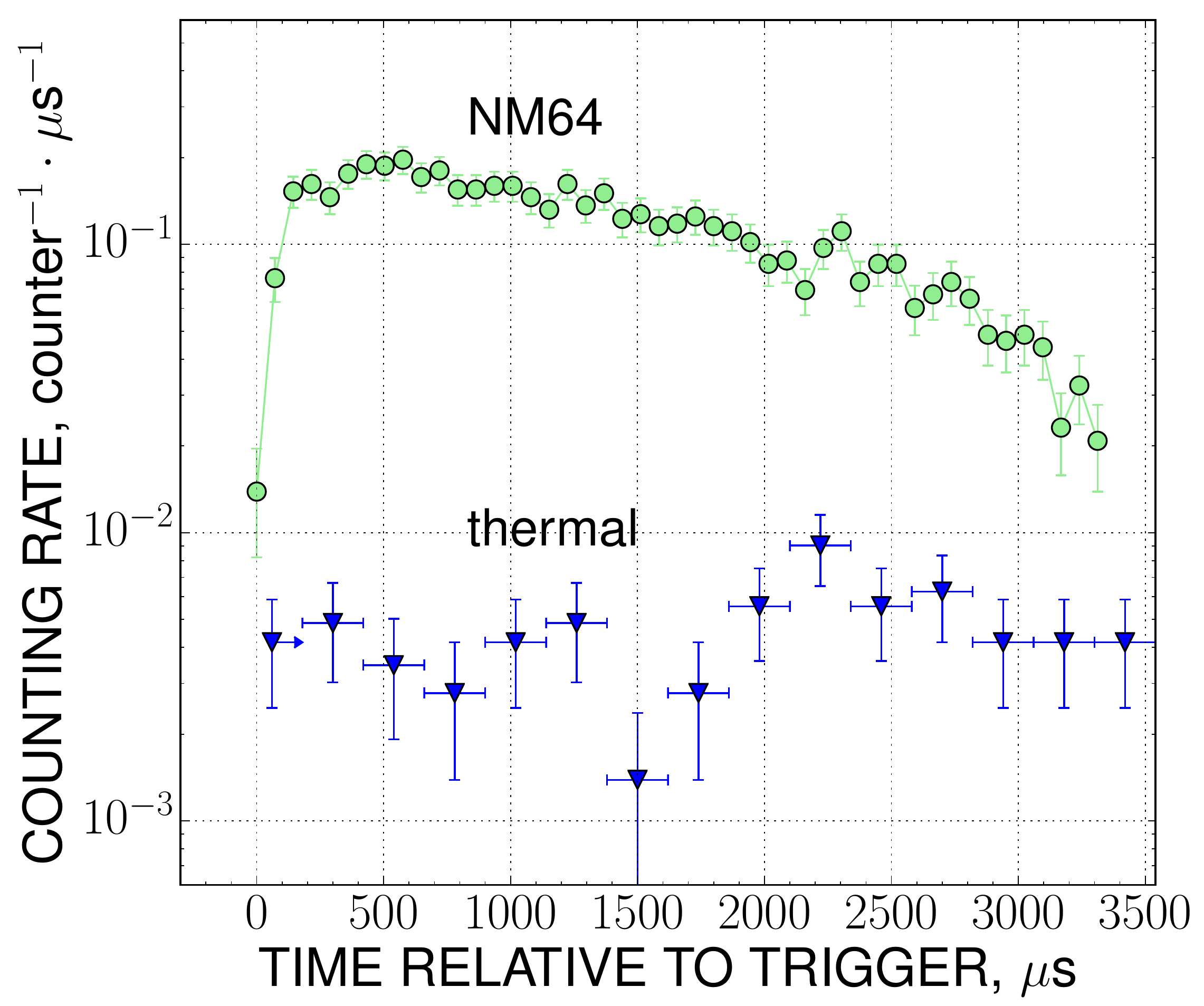}
}
\caption{Time series of neutron intensity registered after the passage of a powerful EAS core ($E_0\approx 10^{17}$eV) immediately through the NM64 supermonitor and a nearby thermal neutron detector.}
\label{figitimi}
\end{center}
\end{figure}

The most characteristic feature in the use of neutron detection method for the study of cosmic ray hadrons is that the energy deposit of hadron CR component is encoded in the {\it number} of the registered pulse signals from proportional neutron counters---the neutron multiplicity $M$. The counting of standard digital pulses exponentially distributed over a prolonged time space of millisecond order due to diffusion of thermalized evaporation neutrons is a simple task from the experimenter's point of view in comparison with any amplitude measurement, which can easily eliminate the influence of saturation effects in a rather wide dynamic range of energy variation, and gives a way to build large-size detector set-ups necessary for investigation of hadronic content in the central core region of high-energy EAS.

Block diagram of the neutron signal acquisition process accepted at the Tien~Shan 18NM64 supermonitor is shown in figure~\ref{figineufuncti}. Analog electric pulses from anode wire of each one of 6~neutron counters in a monitor unit, after proper amplification by a super-high current transfer transistor which resides immediately inside the cap put on the counter's end, are transmitted through short shielded cables to the pulse discriminator board placed in vicinity of this unit. The position of pulse discrimination threshold accepted in neutron detectors is illustrated by the figure~\ref{figineuspecti}; for every counter it was set individually by such a way to occur below the low-amplitude limit of the spectrum of useful neutron pulses, but above the level of random noise of electronic channel. The necessary for this purpose amplitude spectra of pulse signals were measured for all neutron counters in the process of their calibration.

The output pulses from discriminator scheme which have a standardized 3V amplitude and 1$\mu$s duration are connected to the local pulse intensity registration board which operates under the control of the \mbox{STM32F407} type microcontroller \cite{STM32F405}, and is located also in close vicinity to supermonitor (0.5--2m). Each supermonitor unit has its own pair of discriminator/microcontroller boards with 6~informational channels inside.

The embedded program code of the microcontroller ensures complete operation of the input pulse signals from the neutron detectors of the corresponding monitor unit:
\begin{itemize}
{\item the measurement of high-resolution time series of the intensity of neutron signals like the ones shown in figure~\ref{figitimi}: the EAS triggered counting of the number of input pulses registered in every informational channel separately during a number of short (typically, $30-70$$\mu$s) successive time intervals, with their total duration of about $3000-7000$$\mu$s, i.e. of $5-10$~neutron lifetimes in a NM64 supermonitor unit;}
{\item synchronization of the registered time series with an EAS passage which is defined by the {\it external} trigger signal from scintillation shower system, so the initial interval of time sequence is strictly bound to the moment of trigger arrival, and the signal intensity data are available both before and after this moment;}
{\item generation of {\it internal} trigger signal for alternative (and independent on any external condition) synchronization of the time series measurement at the moment when the transient sum multiplicity of input signals which is calculated continuously over some short time intervals after registration of every next input pulse occurs above a predefined threshold value;}
{\item continuous measurement of the counting rate of neutron signals in each informational channel with a low ($10-60$s) time resolution, synchronized only with astronomical time clock, and without any special binding to some trigger for the purpose of steady monitoring of the CR intensity variations.}
\end{itemize}
Due to principally asynchronous organization of the embedded program code, all mentioned tasks are solved simultaneously by the microcontroller unit. The results obtained in these measurements are transmitted through its built-in UART interface and an UART-USB converter to central computer, and later these data are loaded into the corresponding tables of the common database.

All significant operation parameters of the embedded program such as the number and disposition of input informational channels to be traced, the amount and duration of time intervals for the intensity series measurements, and the type of triggering currently used can be set at any time when necessary by corresponding commands from the central control computer which come to autonomous registration board through the same UART interface.

The signal acquisition in low-threshold neutron detectors of the kind shown in figure~\ref{figineudete} is organized in the same way; in this case the dis\-cri\-mi\-na\-tor/mi\-cro\-cont\-rol\-ler boards are placed just inside the box with neutron counters, and the results of the measurements are transmitted to central computer via the sequential RS232 interface line.

\paragraph{The underground neutron detectors}

\begin{figure}
\begin{center}
\includegraphics[width=0.5\textwidth, angle=0, clip,trim=0mm 60mm 50mm 9mm] {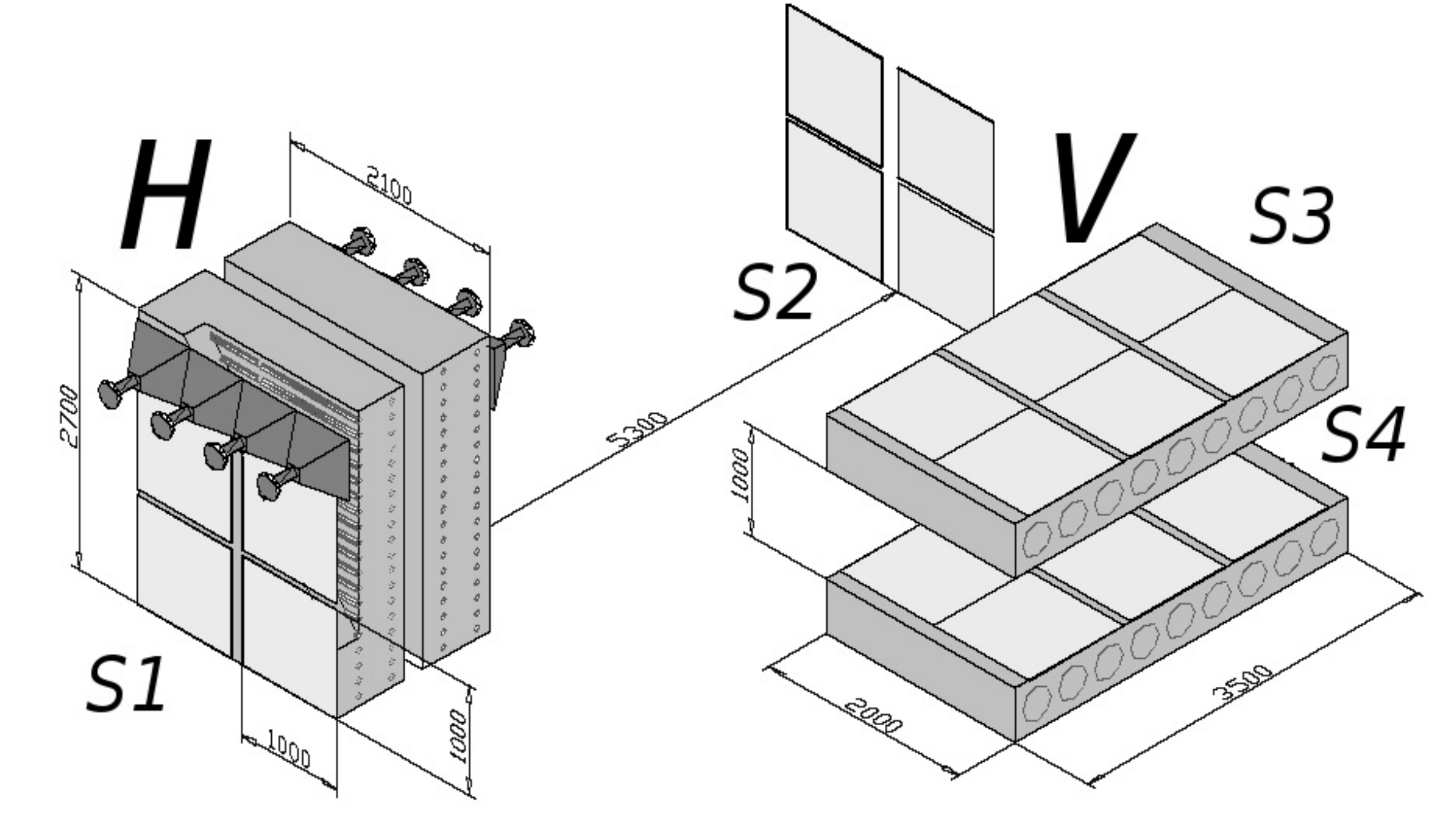}
\end{center}
\caption {Neutron detectors in the underground room of Tien~Shan station. {\it H} and {\it V} -- the horizontal and vertical neutron calorimeters; {\it S1--S4} -- the scintillation planes of two coincidence telescopes. Dimensions are shown in millimeters.}
\label{figundg1}
\end{figure}

\begin{figure*}
\begin{center}
\includegraphics[height=0.7\textwidth, angle=270, clip, trim=0mm 10mm 0mm 20mm] {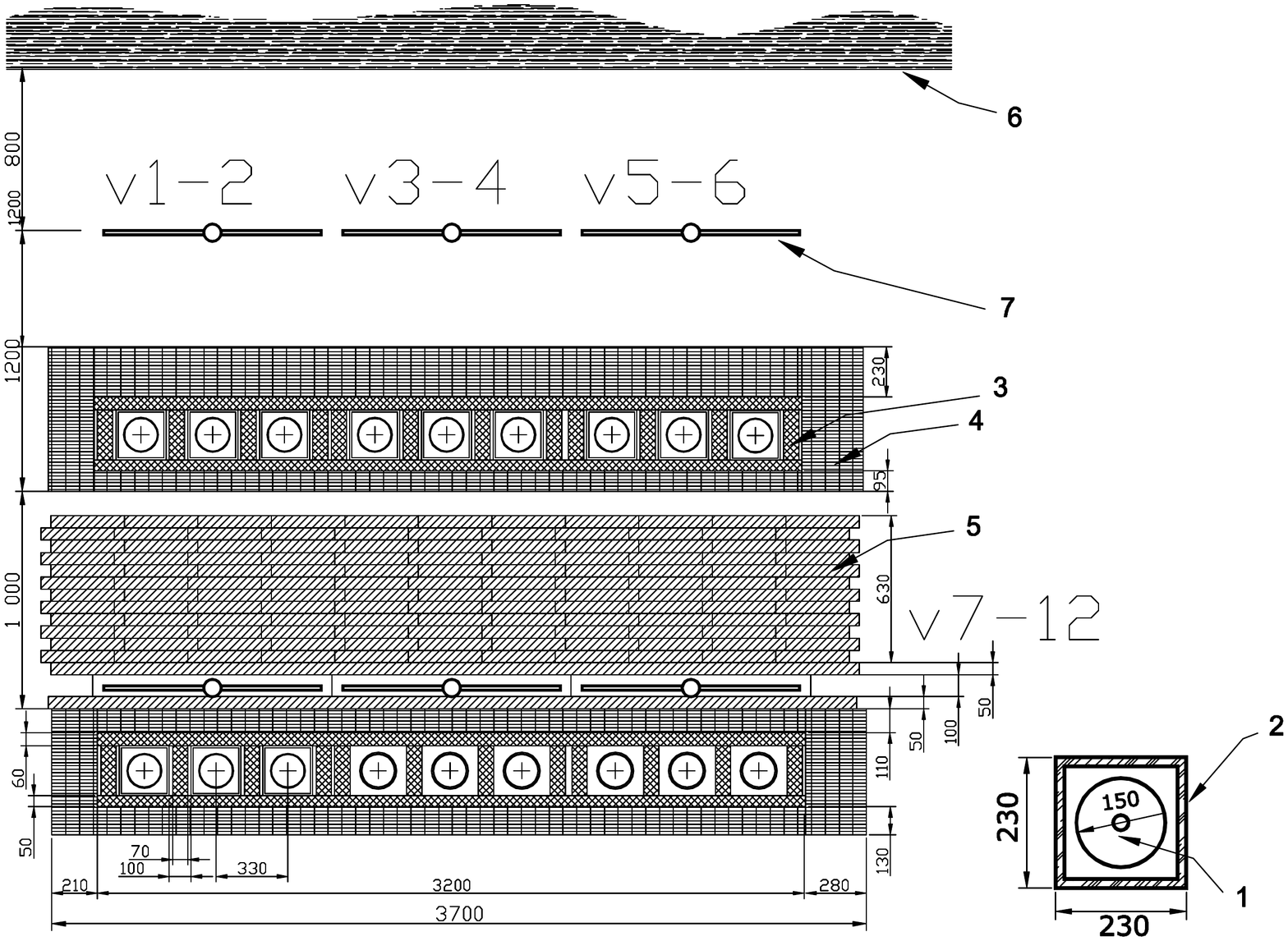}
\\
\includegraphics[height=0.7\textwidth, angle=270, clip, trim=0mm 10mm 0mm 20mm] {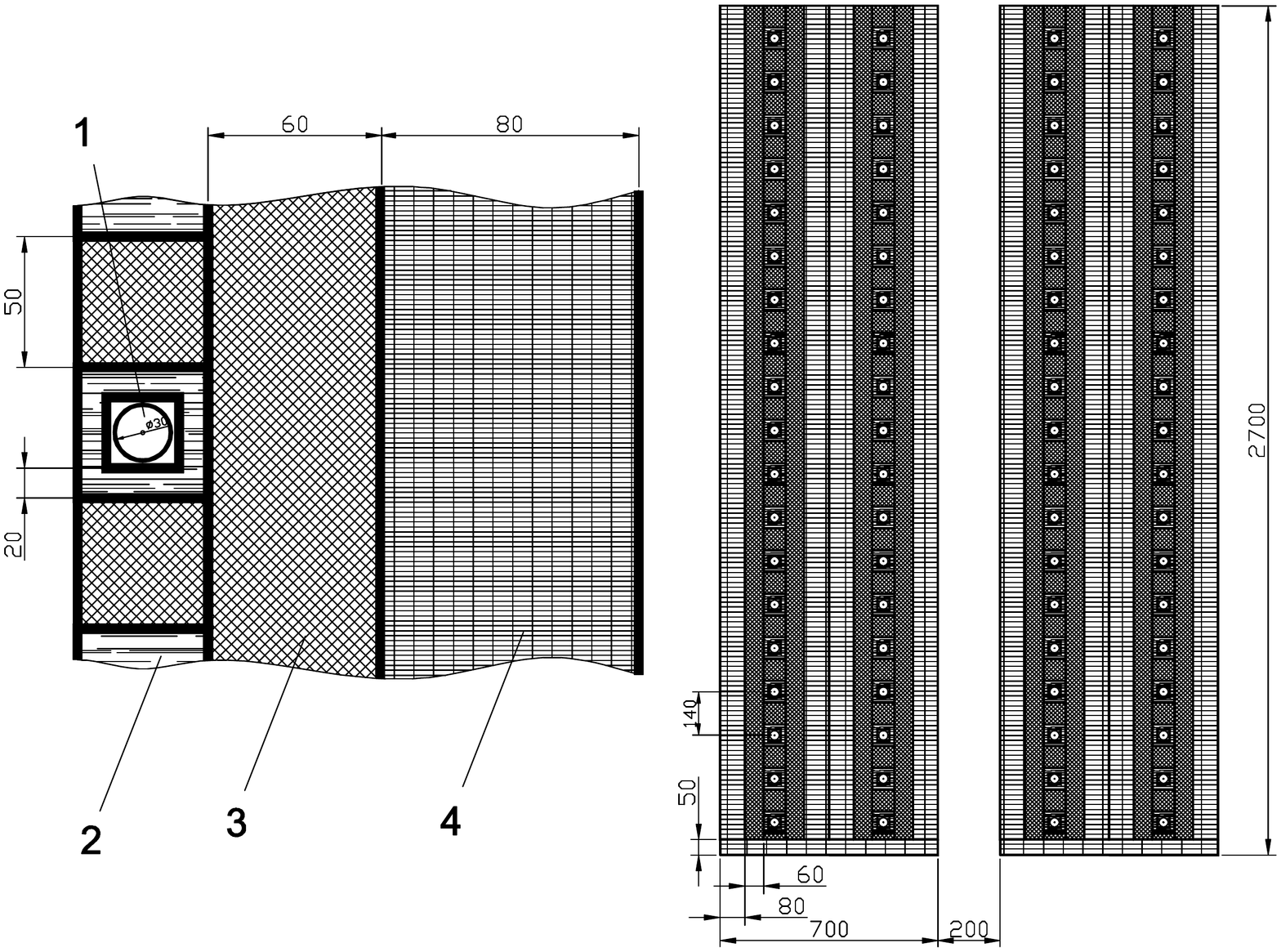}
\end{center}
\caption {Internal arrangement of the horizontal (top) and vertical neutron detectors in the underground room. {\it 1} -- neutron counter; {\it 2} -- wooden box; {\it 3} -- lead; {\it4} -- rubber; {\it 5} -- iron; {\it 6} -- the 2000g/cm$^2$ thick soil layer; {\it 7} -- scintillation detectors {\it v1--v6} and {\it v7--v12} of the telescope coincidence system. Dimensions are shown in millimeters.}
\label{figundg2}
\end{figure*}

The development of the large-sized neutron detector complex in the underground room of the Tien~Shan mountain station was inspired by former observation there of numerous neutron events whose intensity seems to be unexpectedly high in comparison with what should be expected from the nuclear interaction of cosmic ray muons, and which nature still remains unclear \cite{undgour1,undgour2}. Originally, the newly planned set of underground neutron detectors was constructed mainly for the study of this phenomenon.

As it is shown in figure~\ref{figundg1}, at present the underground installation consists of two basic parts: the vertical and the horizontal neutron calorimeters, both built {\it ad exemplum} of the NM64 supermonitor but instead of a single row of neutron counters include just a pair of succeeding detector units with thick lead absorbers inside. The units are placed one after another in the horizontal set-up, and one above the other in the vertical one, so the sum absorber thickness occurs about the length of nuclear interaction of cosmic ray hadrons in each detector. The sum mass of lead absorber used underground is about 35t in horizontal and about 20t in vertical calorimeters.

Successive arrangement of the neutron detector units in the underground room was designed in order to study more precisely the penetrative properties of the cosmic ray component responsible for generating neutron events underground, and to compare its angular distribution with that of the high-energy muons. In supposition that this component may consist of charged particles, both neutron detectors are surrounded with telescope like set-ups built on the basis of large plain scintillation detectors (of the type shown in figure~\ref{figscinti}) which can generate a coincidence trigger to initiate the process of signal recording by the passage of a CR particle through selected diapason of spatial angles, either in nearly vertical or horizontal direction.

Internals of both underground neutron detectors are shown in figure~\ref{figundg2}. The vertical neutron calorimeter consists of a pair of separate units with 9~big boron ionization neutron counters inside, of the same type as the ones used in NM64~supermonitor. Internal arrangement of these units is also made as similar as possible to construction of the standard supermonitor to facilitate the mutual comparison of the data obtained at both installations. As an outer moderator of evaporation neutrons and a reflector of low-energy neutrons which come from outside, in the underground detector are used the sheets of rubber ((CH$_2$)$_n$); and as an inner hydrogen-reach organic moderator surrounding the neutron counters either the polyethylene tubes (as in NM64), or the boxes of wood. Within the empty space which separates both units of vertical calorimeter, an additional mass of different absorber substances can be placed for the investigation of penetrative properties of the neutron bearing CR component; at the present, this gap is completely filled with iron filter having a sum thickness of about 600g/cm$^2$ (which is comparable to the full atmosphere thickness above the Tien~Shan station, 690g/cm$^2$, but is compactly concentrated within a short geometrical distance).

Horizontal detector is made on the basis of $^3$He-filled ionization neutron counters of the same type as in the above described detectors of low-energy neutrons. The counters are grouped in four succeeding vertical planes, by 19 rows of 2 counters (placed one after another) in every plane. Each counter is put inside a wooden box, whose walls play the role of a neutron moderator, and is surrounded by a thick lead absorber, where the evaporation neutrons are to be born in interactions of cosmic ray particles. From outside, the lead assemblage is covered by rubber layers which serve as moderator of evaporation neutrons, and simultaneously as a shielding against the external low-energy neutron background.

Data registration procedure accepted at both underground installations is the same as the one described above for the on-ground neutron detectors. It is anticipated, that the registration process of the time series of signal intensity from all neutron counters placed underground can be triggered with different types of control signals: a scintillation coincidence from the passage of a charged CR particle through the horizontal or vertical pair of telescopic planes; an internal neutron trigger generated at the moment of a transient intensity increase when the current sum number of neutron pulses occurs above some predefined threshold; and by an external EAS trigger signal transmitted from the surface shower installation.

\begin{figure}
\begin{center}
\includegraphics[width=0.48\textwidth, trim=0mm 0mm 0mm 0mm] {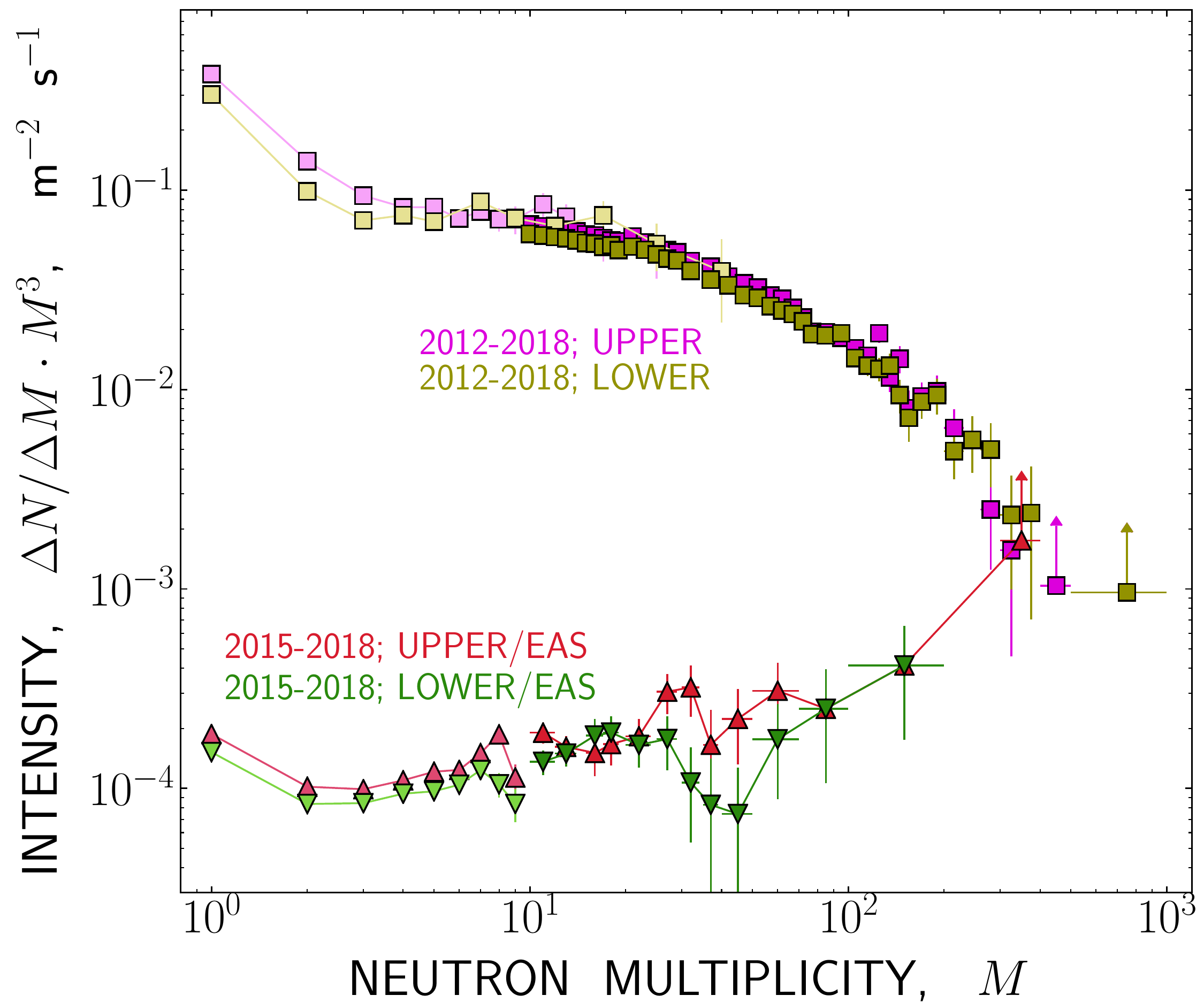}
\end{center}
\caption {The neutron multiplicity spectra of the events registered within the upper and lower units of the vertical underground calorimeter. For the reference, the upper axis corresponds to estimations of the hadron energy deposit in the neutron detector unit which have been calculated according to the neutron monitor calibration formula taken from \cite{nmn2003}: $E_h=0.49\cdot M^{1.8}$GeV.}
\label{figundgspc}
\end{figure}

At the present time the vertical neutron calorimeter in underground room of Tien~Shan station is kept in a ready-to-use state, and in fact it has been operating continuously over the most part of the last decade. The horizontal neutron detector and the scintillation telescopic set-ups around both installations are mounted at their operation places, and the tuning of their data registration channels is now in progress.

As an illustration of the data newly obtained with vertical neutron calorimeter in the underground room, figure~\ref{figundgspc} presents the spectra of neutron events over the multiplicity of registered evaporation neutrons $M$ (which is connected with total energy deposit $E_h$ of initial CR particle inside the detector unit, as it was discussed above). The top pair of spectra shown in this plot were calculated for the events registered either in the upper or in the lower unit of the vertical neutron calorimeter under the control of corresponding internal trigger which has been generated in the moments of fast transient increase of neutron multiplicity inside the unit; generally, these spectra are similar to old ones published in \cite{undgour1}, \cite{undgour2}. Another spectra pair with much lesser absolute intensity which is shown at the bottom of figure~\ref{figundgspc} corresponds to the events triggered by EAS passages in the time of test exploitation of the new shower system in 2015-2016. With increase of statistics, the data of such kind will be useful to clarify the nature of underground neutron events.

\section{The hybrid ionization-neutron calorimeter}

\begin{figure*}
\begin{center}
\includegraphics[width=\textwidth, clip,trim=0mm 0mm 19mm 200mm] {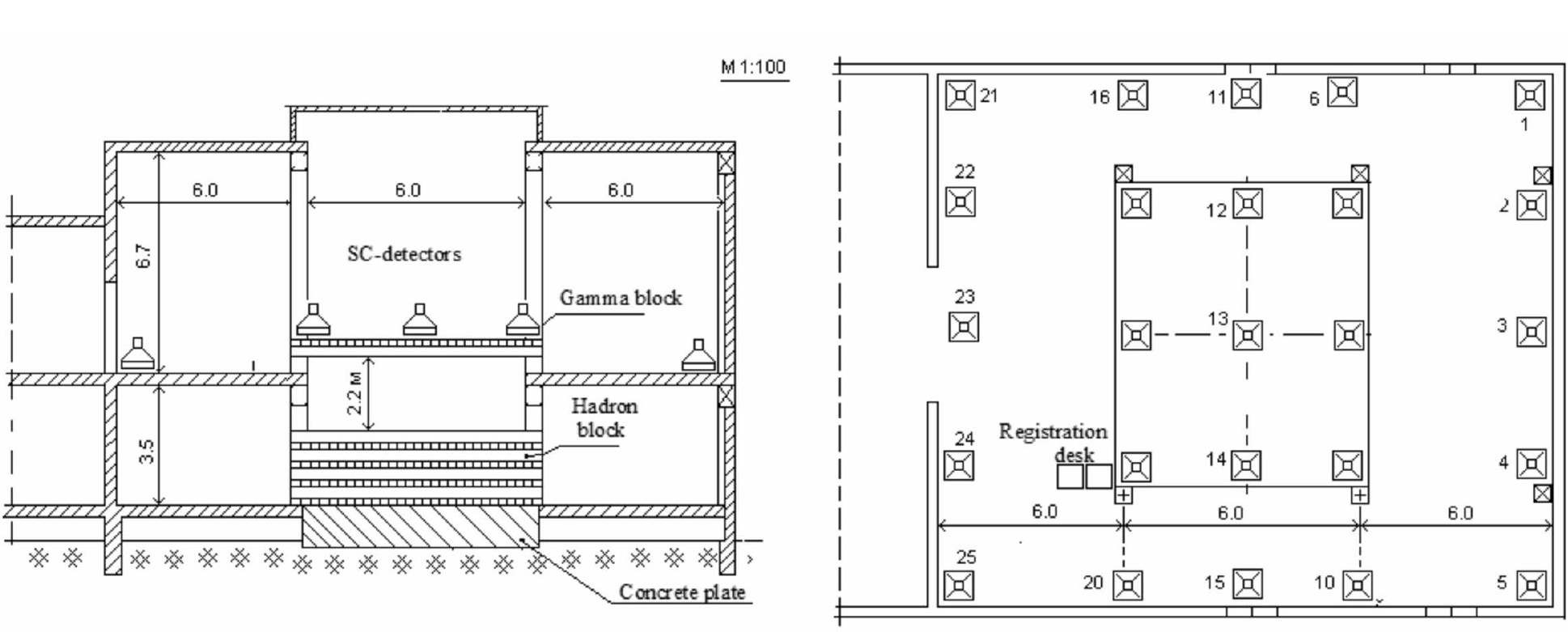}
\\
\includegraphics[width=0.7\textwidth, clip,trim=0mm 0mm 0mm 50mm] {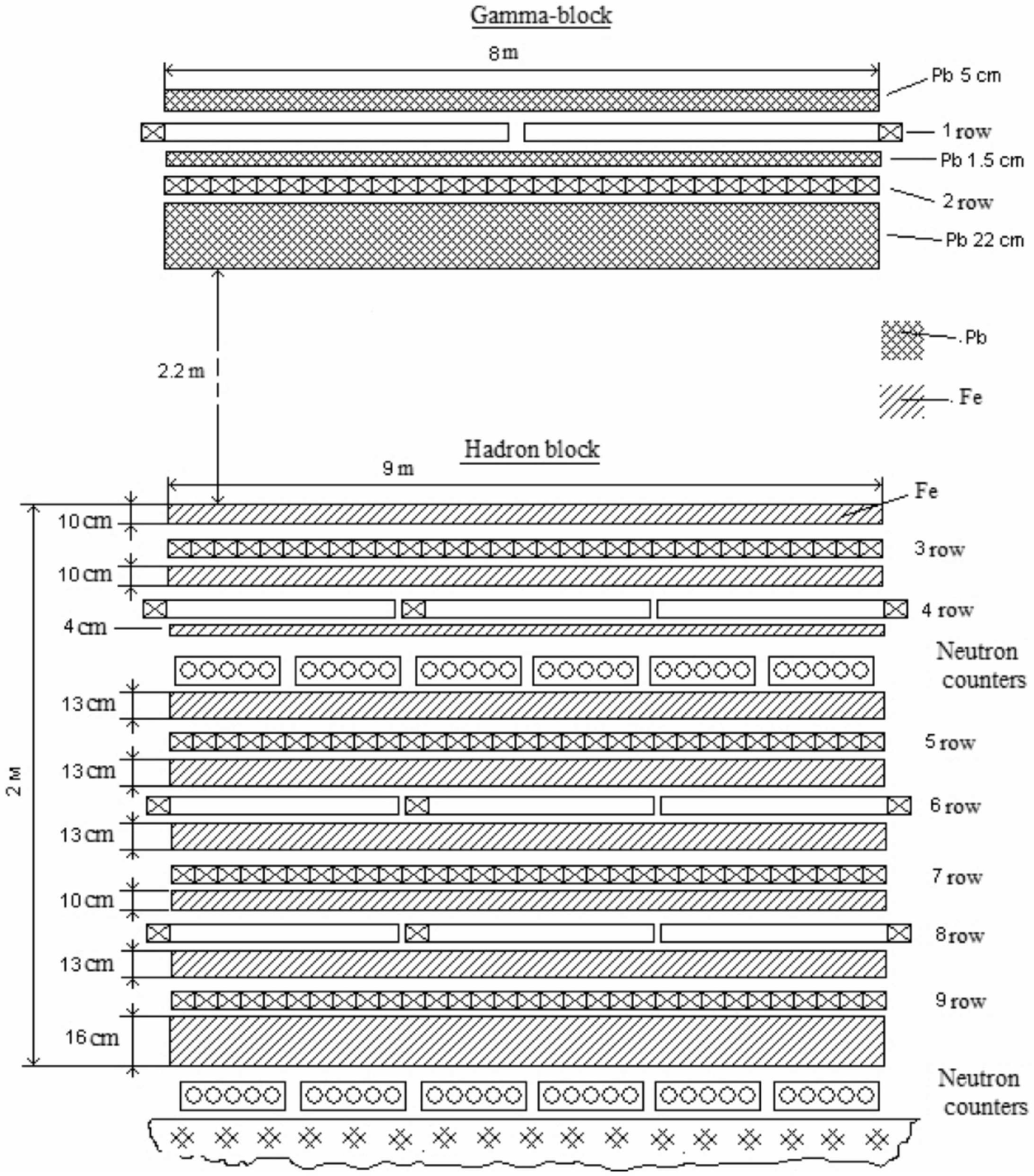}
\end{center}
\caption {Schematics of principal set-up of the {\it Center~II} detector complex (above; all sizes are shown in meters), and internal arrangement of IN\-CA calorimeter.}
\label{figioca}
\end{figure*}

The detector complex {\it Center~II} is placed within and around the capital two-storeyed building with a 324m$^2$ of inner area, and 11m of the height. As a whole, its principal set-up is illustrated by figure~\ref{figioca}.

The central part of the detector complex consists of the hybrid ionization-neutron calorimeter (IN\-CA) which is designed for precise registration of the hadronic CR component, particularly within the EAS core region, and of the local scintillation detector system necessary for selection of EAS which pass in vicinity to the calorimeter. The total sensitive area of IN\-CA is 55m$^2$, a carpet of 25 scintillation detectors ordered equidistantly with a 5m spatial step is situated just above it, and another 10 peripheral scintillators are spread around the calorimeter building at the distances about 10-20m. In local shower subsystem of the {\it Center~II} point the scintillation detectors of both types shown in figure~\ref{figscinti} are applied.

As it is illustrated by the upper drawing of figure~\ref{figioca}, the IN\-CA calorimeter has a double-tiered structure consisting of a pair of separate parts; the upper $\Gamma$-block placed at the top storey of the building, and the deep lower H-block. 

The $\Gamma$-block is intended mostly for registration of the high-energy gamma-radiation and electrons in the central region of an EAS core. As it is shown in the lower picture of figure~\ref{figioca}, it includes two perpendicular rows of ionization chambers interlaced with lead layers of total thickness about 12~cascade units where high-energy electromagnetic cascades are to be developing in. Every row consists of two halves, the ionization chambers in both being oppositely oriented with their backs turned towards each other, so that the whole calorimeter block is logically subdivided into four lesser parts: {\it front}, {\it right}, {\it back}, and {\it left}. Mutually perpendicular orientation of the chamber axes in succeeding rows permits to define the position of electromagnetic cascade in a two-dimensional coordinate frame $(X,Y)$ with precision of the order of 5-10cm according to distribution of secondary ionization signal, and additional subdivision of both rows makes this process somewhat easier in the case of complex multi-cascade events.

Just above the $\Gamma$-block it is also anticipated an arrangement of a standard X-ray emulsion chamber of the same type as the ones which have been used earlier in Pamir \cite{xrec_trudy154} and Hadron \cite{ontienhistory} experiments to ensure a qualitatively higher spatial resolution of the sub-millimeter order.

Immediately beneath the $\Gamma$-block it is located a massive 22cm ($\sim$300g/cm$^2$) thick lead target where the total absorption of the EAS core electromagnetic component must have place, and the high-energy cosmic ray hadrons are to be interacting in; later on, empty 2.2m high vertical air gap is present for decay of short-living mesons which have been originating as the result of these interactions.

The deep hadron calorimeter part residing just below the considered one (H-block) has a massive absorber with total depth about $\sim$600g/cm$^2$ which corresponds to 6-7~interaction length units of CR hadrons. As shown in figure~\ref{figioca}, the H-block from its top to bottom includes 9 successive ionization chamber layers interleaved with thick sheets of iron absorber. In contrast to gamma-block, there is a third {\it middle} sub-row of ionization  chambers in every X-projection layer of hadron block (with chamber axes directed in parallel to the long side of the calorimeter). A large depth of absorber inside the H-block permits to trace precisely the cascade curves originated by CR hadron interaction, and to estimate initial hadron energy with an accuracy of 20-25\% as an integral of ionization signal below the curve.

Because of a large number of vertically spaced detector layers within the IN\-CA system, and a relatively high number of separate registration channels in each layer a possibility should be mentioned here to use the described calorimetric set-up as an alternative mean to define the EAS arrival angles by approximation of the position of EAS core imprints in succeeding rows of ionization chambers. It is obvious that this possibility is fully independent and complementary one to the traditional procedure of the time delay measurements between scintillation signals which is accepted for this purpose at the {\it Center} shower system, and presents another informational channel for the studies connected with astrophysical aspect of the CR investigation.

The ionization chambers of the IN\-CA calorimeter are made of the 3m long peaces of copper waveguide with a 11$\times$6cm$^2$ rectangular cross-section. The chambers are filled with technical argon gas up to the pressure of 2~atmospheres, and have an inner anode electrode consisting of a continuous $\diameter$3mm metallic tube placed lengthwise in the middle of each chamber along its axis. The operational feeding voltage of 600V is connected immediately to the anode.

\begin{figure*}
\begin{center}
\includegraphics[width=1\textwidth, clip,trim=10mm 40mm 10mm 40mm] {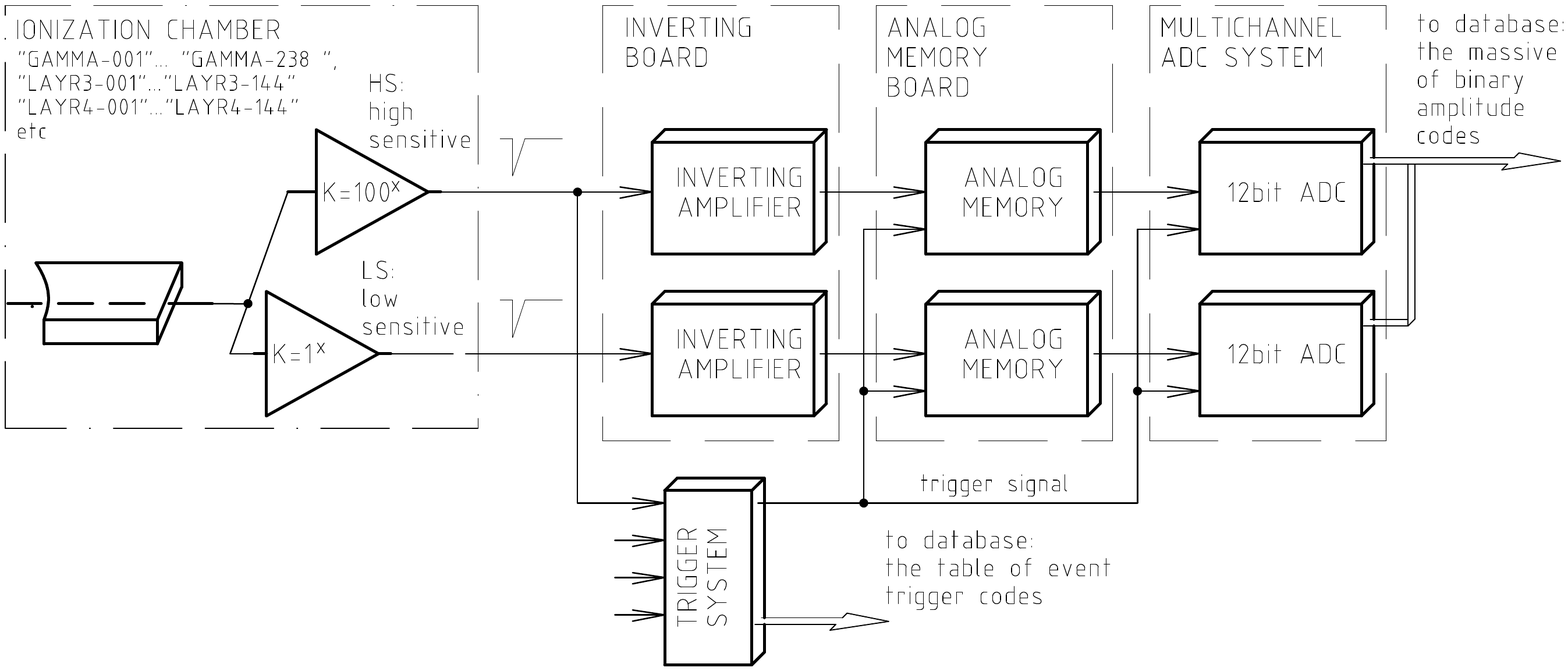}
\end{center}
\caption {Block diagram of an ionization chamber data acquisition channel.}
\label{figiocadaq}
\end{figure*}

A useful physical signal of the ionization chamber is the amplitude voltage of the electric pulse at its anode output which is registered by multichannel ADC system with some delay after an EAS passage. The whole acquisition sequence of the ionization chamber signals is presented in figure~\ref{figiocadaq}; mainly, it is quite analogous to corresponding scheme of the scintillation shower system. In contrast to the latter, only two amplitude diapasons, high- ({\it HS}) and low-sensitive ({\it LS}) ones, are anticipated for the use in the ionization channels; as it was shown in section~\ref{sedaq} such an arrangement can ensure a dynamic range of amplitude measurements about $(5-7)\cdot10^4$. The signals of scintillation carpet, which is placed above and around the calorimeter, are operated completely in the way described in sections~\ref{sesci} and~\ref{sedaq}.

It is planned that the trigger signal to control the IN\-CA ADC system could be generated by the local shower trigger unit of {\it Center~II} registration point which works by summing of {\it MS} diapason pulses from scintillation carpet placed above the $\Gamma$-block, quite analogously to the case of central shower system which was considered in section~\ref{sesci}. Also, the {\it HS} diapason output signals of ionization chambers in succeeding calorimeter layers are summed together in various combinations ({\it front}/{\it right}, {\it back}/{\it right} and the like) to form the local IN\-CA trigger necessary for registration of primary CR hadrons which have ''slipped'' the entire atmosphere above the calorimeter without generation of any noticeable EAS. For registration of large high-energy EAS the control of the whole calorimeter system by the outer shower trigger from its counterpart {\it Center} point is also possible.

Besides the ionization chamber method which is based on the measurement of electromagnetic contents of the secondary electron-photon cascades, the {\it hybrid} IN\-CA project presumes the use of neutron detectors for registration of evaporation neutrons born by high-energy interaction of cosmic ray hadrons with heavy absorber nuclei within the calorimeter. As the results of primary investigations \cite{inca2006,inca2007,inca2008} have shown, the informativeness of such a combined calorimeter type is substantially higher in comparison with that of ionization and neutron signal detectors used separately. Indeed, in addition to possibility to define the energy deposit of hadronic CR component inside the calorimeter by two independent methods, the neutron yield from the primary gamma-ray induced cascades is known to be an order of magnitude below that of nuclear ones \cite{inca1998c,inca1999c,inca2001}, which circumstance opens a perspective to select the fully electromagnetic EAS born by the high energy CR gamma-quanta, and, again, occurs to be interesting from astrophysical point of view.

There must also be mentioned here once more a wide dynamic diapason and practical operation convenience of the neutron based registration technique of hadronic CR particles which have been discussed in section~\ref{seneu}.

To place the neutron detectors within the IN\-CA calorimetric set-up, a special holes are anticipated below its 4th and 9th rows of ionization chambers (see the figure~\ref{figioca}) where a number of thick-wall wooden boxes with proportional neutron and gamma-ray counters inside are installed (the organic substance these boxes consist of, in particular, plays the role of neutron moderator). The neutron counters applied here for direct neutron registration are of the $^3$He-filled type (see the section~\ref{seneu}); while the ones of gamma-radiation kind are the glass $\diameter 6\times56$cm Geiger-M\"{u}ller tubes with argon filling. The latter ones are intended for the registration of delayed gamma-ray quanta emitted by neutron captures with iron nuclei which take place inside the calorimeter absorber. These counters are placed in proportion "10~gamma + 1~neutron" inside every box to make a single neutron signal registration module. Also, the use of a gadolinium doped oil-paints in coating of counter boxes is planned to enhance the neutron conversion into gamma-ray component.

To register digital pulse signals from both kinds of neutron sensors placed inside the IN\-CA calorimeter it is employed a well-elaborated procedure of the high-re\-so\-lu\-tion time series measurement of pulse intensity which was described in section~\ref{seneu}. For synchronization of the registered time series with the moment of an EAS passage there can be used all types of local IN\-CA trigger (both the trigger from shower scintillation carpet, and the internal hadronic one based on the sum of pulse amplitudes from various combinations of the IN\-CA ionization chambers), so as the outer trigger from the {\it Center} shower system.

\section{Conclusion}

The process of drastic modification of the Tien~Shan mountain cosmic ray installation, which has been going on during the last decades, is now approaching its final stage. At the present time the complex of EAS detectors of Tien~Shan station includes two shower scintillation carpets, the system of timing detectors for determination of the EAS arrival directions, the underground muon and hadron detectors for registration of penetrating shower particles, the hybrid ionization-neutron calorimeter, NM64 type neutron supermonitor, and other neutron detectors for registration of EAS hadronic component in different energy ranges. The full stack of data registration algorithms and operation methods has been elaborated and tested in dealing with real shower events.

The principal design of the newly developed detector complex of the Tien~Shan station is based on simultaneous registration of electromagnetic, muonic and hadron components of the cosmic ray  showers, and special attention is paid to the possibility of precise investigation of the EAS core region. The complex is aimed for investigation of various fundamental and applied problems in different fields of the astroparticle, atmospheric, and environmental physics:
\begin{itemize}
{\item high-precision measurements of the cosmic ray spectrum, mass composition, and distribution of arrival directions in the primary energy range of $E_0\sim 10^{14}-10^{17}$eV with the use of a wide-spread scintillation shower system, a hybrid ionization-neutron calorimeter, muon detectors, and the detectors of Cherenkov radiation \cite{ontiencheren};}
{\item the investigation of high-energy nuclear interaction parameters, especially in the region of EAS core with the use of a dense carpet of scintillation detectors with enhanced up to $(1-2)\cdot 10^6$~dynamic range of amplitude measurements, in combination with ionization calorimeter, saturation-free neutron detectors, X-ray films, underground neutron and muon installations;}
{\item search for the strong interacting dark matter particles whose existing in the Universe is supposed to be in the form of some exotic nuclei (e.g. strangelets) with anomalous combination of low charge $Z\lesssim10$ and large mass $A\gtrsim 350$ according to their specific signature in the complex detector set-up;}
{\item monitoring of the intensity of low-energy $10^{8}-10^{11}$eV cosmic rays with application of various types of neutron and gamma-radiation detectors sensitive to neutron radiation in different ranges of energy spectrum; control of the {\it cosmic weather} \cite{space_weather_inbook}, and of the energetic disturbance events of the solar origin \cite{ongle};}
{\item the study the role of cosmic rays in various environmental phenomena: lightning initiation in the atmosphere \cite{thunderourufn}, intensity variations of the near-Earth gamma-ray background \cite{rainsour2009}, anticipated seismic effects from powerful EAS passage \cite{undgmouacou}, etc.}
\end{itemize}

Test measurements which have been carried out in 2014-2016 confirmed the efficient operation of shower detector subsystems of the Tien~Shan EAS installation. The results obtained so far demonstrate the reliable measurement of the EAS particle density and shower size spectrum within the energy range of $10^{14}-10^{17}$eV. High-resolution time series of neutron intensity, and the spectra of energy deposits into hadron component were registered both on the surface of the Tien~Shan station and underground by the neutron monitor like installations used for the detection of neutron bearing cosmic ray particles. Similar neutron detector technique is planned to be widely applied in operation of the big hybrid Tien~Shan calorimeter IN\-CA.

Current development level of different detector subsystems at the Tien~Shan station allows to expect new results in various fields of the cosmic ray physics in near future.

\section*{Acknowledgements}
This work was supported by the RAS Programs 9P and 10OF and by the projects of Kazakhstan Republic \#0080/GF4 and \#0003-3/PTF-15-AKMIR/0-15.

\section*{References}

\begin{thebibliography}{10}

\expandafter\ifx\csname href\endcsname\relax
  \def\href#1#2{#2} \def\path#1{#1}\fi

\bibitem{2011letessier}
A.~{Letessier-Selvon}, T.~Stanev, Ultrahigh energy cosmic rays, {Rev. Mod.
  Phys.} 83 (2011) 907--942.
\newblock \href {http://dx.doi.org/10.1103/RevModPhys.83.907}
  {\path{doi:10.1103/RevModPhys.83.907}}.

\bibitem{2005kascade}
T.~Antoni, W.~D. Apel, A.~F. Badea, \etal
  {(KASCADE Collaboration)}, {KASCADE} measurements of energy spectra for
  elemental groups of cosmic rays: {R}esults and open problems, {Astropart.
  Phys.} 24 (2005) 1--25.
\newblock \href {http://arxiv.org/abs/[astro-ph/0505413]}
  {\path{arXiv:[astro-ph/0505413]}}, \href
  {http://dx.doi.org/10.1016/j.astropartphys.2005.04.001}
  {\path{doi:10.1016/j.astropartphys.2005.04.001}}.

\bibitem{2001hires}
T.~Abu-Zayyad, K.~Belov, D.~J. Bird, \etal, Measurement of the cosmic-ray energy spectrum and composition from
  $10^{17}$ to $10^{18.3}${eV} using a hybrid technique, {Astrophys. J.} 557
  (2001) 686.
\newblock \href {http://arxiv.org/abs/[astro-ph/0010652]}
  {\path{arXiv:[astro-ph/0010652]}}, \href
  {http://dx.doi.org/10.1016/j.astropartphys.2005.04.001}
  {\path{doi:10.1016/j.astropartphys.2005.04.001}}.

\bibitem{2010auger}
J.~Abraham, P.~Abreu, M.~Aglietta,  \etal {(Pierre
  Auger Collaboration)}, Measurement of the energy spectrum of cosmic rays
  above $10^{18}${eV} using the {Pierre Auger Observatory}, {Phys. Lett. B} 685
  (2010) 239.
\newblock \href {http://arxiv.org/abs/[astro-ph.HE/1002.1975]}
  {\path{arXiv:[astro-ph.HE/1002.1975]}}, \href
  {http://dx.doi.org/10.1016/j.physletb.2010.02.013}
  {\path{doi:10.1016/j.physletb.2010.02.013}}.

\bibitem{2010auger_b}
R.~U. Abbasi, T.~Abu-Zayyad, M.~Allen, First observation of the
  {Greizen-Zatsepin-Kuzmin} supression, {Phys. Rev. Lett.} 100 (2008) 101101.
\newblock \href {http://dx.doi.org/10.1103/PhysRevLett.100.101101}
  {\path{doi:10.1103/PhysRevLett.100.101101}}.

\bibitem{2015telescopearray}
T.~Abu-Zayyad, R.~Aida, M.~Allen,  \etal, Energy spectrum of ultra-high
  energy cosmic rays observed with the {T}elescope {A}rray using a hybrid
  technique, {Astropart. Phys.} 61 (2015) 93--101.
\newblock \href {http://dx.doi.org/10.1016/j.astropartphys.2014.05.002}
  {\path{doi:10.1016/j.astropartphys.2014.05.002}}.

\bibitem{1966gzk_a}
K.~Greisen, End to cosmic-ray spectrum, {Phys. Rev. Lett.} 16 (1966) 748.
\newblock \href {http://dx.doi.org/10.1103/PhysRevLett.16.748}
  {\path{doi:10.1103/PhysRevLett.16.748}}.

\bibitem{1966gzk_b}
G.~T. Zatsepin, V.~A. Kuzmin, Upper limit of spectrum of cosmic rays, {JETP
  Lett.} 4 (1966) 48.

\bibitem{oneasbook}
P.~Grieder, {Extensive Air Showers: High Energy Phenomena and Astrophysical
  Aspects -- A Tutorial, Reference Manual and Data Book}, {Springer Science \&
  Business Media}, Heidelberg Dordrecht London New York, 2010.
\newblock \href {http://dx.doi.org/10.1007/978-3-540-76941-5}
  {\path{doi:10.1007/978-3-540-76941-5}}.

\bibitem{1984akeno}
M.~Nagano, T.~Hara, Y.~Hatano,  \etal, Energy spectrum of primary cosmic rays between
  $10^{14.5}$ and $10^{18}$~{eV}, {J. Phys. G} 10 (1984) 1295--1310.

\bibitem{2008aragats}
A.~P. Garyaka, R.~M. Martirosov, S.~V. Ter-Antonyan, \etal, An all-particle primary
  energy spectrum in the 3-200 {PeV}, {J. Phys. G} 35 (2008) 115201.
\newblock \href {http://dx.doi.org/10.1088/0954-3899/35/11/115201}
  {\path{doi:10.1088/0954-3899/35/11/115201}}.

\bibitem{2008tibet}
M.~Amenomory, X.~J. Bi, D.~Chen, \etal, The
  all-particle spectrum of primary cosmic rays in the wide energy range from
  $10^{14}$ to $10^{17}$~{eV} observed with the {Tibet-III} air-shower array,
  {Astrophys. J.} 678 (2008) 1165--1179.

\bibitem{2012kascade}
W.~D. Apel, J.~C. Arteaga-Velazquez, K.~Bekk, \etal {(KASCADE Collaboration)},
  The spectrum of high-energy cosmic rays measured with {KASCADE-Grande},
  {Astropart. Phys.} 36 (2012) 183--194.
\newblock \href {http://arxiv.org/abs/[astro-ph/1206.3834]}
  {\path{arXiv:[astro-ph/1206.3834]}}, \href
  {http://dx.doi.org/10.1016/j.astropartphys.2012.05.023}
  {\path{doi:10.1016/j.astropartphys.2012.05.023}}.

\bibitem{1994flyseye}
D.~J. Bird, S.~C. Corbato, H.~Y. Dai, \etal, The cosmic-ray energy spectrum observed by the {Fly’s Eye},
  {Astrophys. J.} 424 (1994) 491--502.
\newblock \href {http://dx.doi.org/10.1086/173906} {\path{doi:10.1086/173906}}.

\bibitem{2000hegra}
F.~Arqueros, J.~A. Barrio, K.~Bernlohr, \etal, Energy spectrum and chemical composition of cosmic rays between 0.3
  and 10~{PeV} determined from the {C}herenkov-light and charged-particle
  distributions in air showers, {Astron. Astrophys.} 359 (2000) 682--694.

\bibitem{2012tunka}
S.~F. Berezhnev, D.~Besson, N.~Budnev, \etal, The {Tunka-133} {EAS}
  {C}herenkov light array: status of 2011, {Nucl. Instrum. Methods A} 692
  (2012) 98--105.
\newblock \href {http://arxiv.org/abs/[astro-ph.HE/1201.2122]}
  {\path{arXiv:[astro-ph.HE/1201.2122]}}, \href
  {http://dx.doi.org/10.1016/j.nima.2011.12.091}
  {\path{doi:10.1016/j.nima.2011.12.091}}.

\bibitem{2015yakutsk}
S.~Knurenko, I.~Petrov, Z.~Petrov, I.~Sleptsov, Ultra-high energy cosmic rays:
  40 years retrospective of continuous observations at the {Yakutsk} array:
  {P}art 1. {C}osmic ray spectrum in the energy range $10^{15}-10^{18}${eV} and
  its interpretation, {EPJ Web Conf.} 99 (2015) 04001.
\newblock \href {http://dx.doi.org/10.1051/epjconf/20159904001}
  {\path{doi:10.1051/epjconf/20159904001}}.

\bibitem{scinti1}
T.~P. Amineva, V.~S. Aseykin, Y.~N. Vavilov, \etal,  The Installation for
  Study of Extensive Air Showers and Nuclear Interactions of the Cosmic Ray
  Particles with the Energy $10^{12}-10^{16}$~{eV}, {Nauka, Moscow}, 1970, pp.
  157--176 {(in Russian)}.

\bibitem{xrec_trudy154}
{PAMIR Collaboration}, Investigation of the nuclear interactions in cosmic rays
  in the energy range of $10^{14}-10^{17}$eV with the method of X-ray emulsion
  chambers ({'Pamir'} experiment), {Nauka, Moscow}, 1984, pp. 3--141 {(in
  Russian)}.

\bibitem{2008chacaltaya}
C.~Aguirre, H.~Aoki, K.~Hashimoto, \etal, Study of hadronic component in air showers
  at {Mt. Chacaltaya}, {Nucl. Phys. B (Proc. Suppl.)} 75 (1999) 186--190.

\bibitem{ontienhistory}
S.~F. Abdrashitov, D.~S. Adamov, V.~V. Arabkin, \etal, The {Adron} apparatus
  for researching the primary cosmic radiation and the characteristics of
  nuclear interactions in the atmosphere, {Bull. Acad. Sci. USSR Phys. Ser.} 50
  (1986) 123--126.

\bibitem{2001auger}
J.~J. Beatty, { for Pierre Auger Collaboration}, The {Pierre Auger} project:
  {A}n observatory for the highest energy cosmic rays, {Int. J. Mod. Phys. A}
  16 (2001) 1022.
\newblock \href {http://dx.doi.org/10.1142/S0217751X01008771}
  {\path{doi:10.1142/S0217751X01008771}}.

\bibitem{2008icetop}
T.~Waldenmaier, {IceTop -- C}osmic ray physics with {IceCube}, {Nucl. Instrum.
  Methods A} 588 (2008) 130--134.
\newblock \href {http://arxiv.org/abs/[astro-ph.HE/0802.2540]}
  {\path{arXiv:[astro-ph.HE/0802.2540]}}, \href
  {http://dx.doi.org/10.1016/j.nima.2008.01.015}
  {\path{doi:10.1016/j.nima.2008.01.015}}.

\bibitem{2012telescopearray}
T.~Abu-Zayyad, R.~Aida, M.~Allen, \etal, The surface detector array of the
  {T}elescope {A}rray experiment, {Nucl. Instrum. Methods A} 689 (2012) 87--97.
\newblock \href {http://dx.doi.org/10.1016/j.nima.2012.05.079}
  {\path{doi:10.1016/j.nima.2012.05.079}}.

\bibitem{ontienvilson}
V.~V. Guseva, E.~V. Denisov, N.~A. Dobrotin, \etal, New installation for the
  study of strong interaction at 100-1000{GeV} of the {Tien~Shan} mountain
  station, {Bull. Acad. Sci. USSR Phys. Ser.} 30 (1966) 1574--1576 {(in
  Russian)}.

\bibitem{ontiensparkchamber}
K.~A. Kotel'nikov, Y.~E. Zvonkov, S.~B. Shaulov, Double-gap 6m$^2$ spark
  chamber, {Instrum. Exp. Tech. (USSR)} 1 (1977) 66--67 {(in Russian)}.

\bibitem{longflying2}
V.~I. Yakovlev, Long flying component: recent data and interpretation, in:
  {International Symposium on VHE CRI}, Ann Arbor, USA, 1992, p. 154.

\bibitem{longflying3}
I.~M. Dremin, \etal, {Monte Carlo} simulations of long-flying cascades in
  cosmic rays and leading charm at {SSC}, in: {International Symposium on VHE
  CRI}, Ann Arbor, USA, 1992, p. 534.

\bibitem{longflying_pamir}
V.~S. Hlytchieva, L.~G. Sveshnikova, \etal, Anomalies of hadron attenuation
  at large depth of lead, in: {Proceedings of the VI ISVHECRI}, Tarbes, France,
  1990, p. 184.

\bibitem{longflying_pamir2}
{PAMIR Collaboration}, Observation of attenuation behavior of hadrons in
  extremely high energy cosmic ray interactions: {N}ew hadronic state?, {Nucl.
  Phys. B} 424 (1994) 241--287.
\newblock \href {http://dx.doi.org/10.1016/0550-3213(94)90295-X}
  {\path{doi:10.1016/0550-3213(94)90295-X}}.

\bibitem{halo1}
V.~V. Arabkin, V.~A. Borodkin, M.~D. Smirnova, \etal,
  The haloes spectra of super families at the pressures 585g/sm$^2$ and
  690g/sm$^2$ and nucleons absorption path at the energy about $10^{16}${eV},
  in: {Proceedings of the VI ISVHECRI}, Tarbes, France, 1990, pp. 257--260.

\bibitem{halo2}
{PAMIR Collaboration}, Halo development in deep lead {X}-ray emulsion chamber,
  {Bull. Russ. Acad. Sci. Phys.} 55 (1991) 33--37.

\bibitem{halo3}
H.~Aoki, K.~Hashimoto, K.~Honda, \etal, \href{http://stacks.iop.org/0954-3899/30/i=2/a=012}{A
  halo event observed with an emulsion chamber and air shower array at {Mt
  Chacaltaya}}, {J. Phys. G} 30~(2) (2004) 137.

\bibitem{alignment1}
R.~A. Mukhamedshin, \href{http://stacks.iop.org/1126-6708/2005/i=05/a=049}{On
  coplanarity of most energetic cores in gamma-ray–hadron families and hadron
  interactions at $\sqrt{s}\ge 4${TeV}}, {J. High Energy Phys.} 2005 (2005)
  049.
\newblock \href {http://dx.doi.org/10.1088/1126-6708/2005/05/049}
  {\path{doi:10.1088/1126-6708/2005/05/049}}.

\bibitem{alignment2}
V.~V. Kopenkin, A.~K. Managadze, I.~V. Rakobolskaya, T.~M. Roganova, Alignment
  in gamma-hadron families of cosmic rays, {Phys. Rev. D} 52 (1995) 2766.
\newblock \href {http://arxiv.org/abs/[hep-ph/9408247]}
  {\path{arXiv:[hep-ph/9408247]}}, \href
  {http://dx.doi.org/10.1103/PhysRevD.52.2766}
  {\path{doi:10.1103/PhysRevD.52.2766}}.

\bibitem{alignment3}
M.~C. Talai, R.~Attallah, J.~Capdevielle, Aligned events observed by emulsion
  chambers in the knee region, {Int. J. Mod. Phys. A} 20 (2005) 6849.
\newblock \href {http://dx.doi.org/10.1142/S0217751X05030284}
  {\path{doi:10.1142/S0217751X05030284}}.

\bibitem{hadron_high_muons}
S.~B. Shaulov, S.~P. Bezshapov, Looking for strange quark matter in cosmic
  rays, {The Eur. Phys. J. Conf.} 52 (2013) 04010.
\newblock \href {http://dx.doi.org/10.1051/epjconf/20125204010}
  {\path{doi:10.1051/epjconf/20125204010}}.

\bibitem{hadron_scaling}
V.~V. Arabkin, Z.~Zhanseitova, K.~V. Cherdyntseva, \etal, {Anomalous behavior of gamma family characteristics with a
  primary energy of $10^{16}$eV}, {Bull. Russ. Acad. Sci. Phys.} 55N4 (1991)
  51--54, [Izv. Ross. Akad. Nauk Ser. Fiz.55,674(1991)].

\bibitem{jopg2001}
V.~P. Antonova, A.~P. Chubenko, S.~V. Kryukov, \etal, Anomalous time structure of
  extensive air shower particle flows in the knee region of primary cosmic ray
  spectrum, {J. Phys. G} 28~(2) (2002) 251--266.
\newblock \href {http://dx.doi.org/10.1088/0954-3899/28/2/306}
  {\path{doi:10.1088/0954-3899/28/2/306}}.

\bibitem{jopg2008}
A.~Chubenko, A.~Shepetov, V.~Antonova, \etal, The influence
  of background radiation on the events registered in a neutron monitor at
  mountain heights, {J. Phys. G} 35 (2008) 085202.
\newblock \href {http://dx.doi.org/10.1088/0954-3899/28/2/306}
  {\path{doi:10.1088/0954-3899/28/2/306}}.

\bibitem{undgour1}
A.~P. Chubenko, A.~L. Shepetov, L.~I. Vildanova, \etal, Neutron events in the underground monitor of the {Tien Shan}
  high-altitude station, {Bull. Lebedev Phys. Inst.} 34~(4) (2007) 107--113.
\newblock \href {http://dx.doi.org/10.3103/S1068335607040033}
  {\path{doi:10.3103/S1068335607040033}}.

\bibitem{undgour2}
A.~P. Chubenko, A.~L. Shepetov, V.~V. Oscomov, \etal, The underground
  neutron events at {Tien-Shan}, in: {Proceedings of the 30th ICRC}, Vol. 4 (HE
  part 1), M\'{e}xico City, M\'{e}xico, 2008, pp. 3--6.

\bibitem{centauro0}
C.~M.~G. Lattes, Y.~Fujimoto, S.~Hasegawa, Hadronic interactions of high energy
  cosmic-ray observed by emulsion chambers, {Phys. Rep.} 65 (1980) 151.

\bibitem{centauro2}
{PAMIR Collaboration}, Observation of a high-energy cosmic-ray family caused by
  a {C}entauro-type nuclear interaction in the joint emulsion chamber
  experiment at the {Pamirs}, {Phys. Lett. B} 190 (1987) 226--233.
\newblock \href {http://dx.doi.org/10.1016/0370-2693(87)90871-9}
  {\path{doi:10.1016/0370-2693(87)90871-9}}.

\bibitem{centauro1}
{JACEE Collaboration}, {JACEE} results on very high energy interactions, {Nucl.
  Phys. B (Proc. Suppl.)} 52 (1997) 81--91.
\newblock \href {http://dx.doi.org/10.1016/S0920-5632(96)00851-1}
  {\path{doi:10.1016/S0920-5632(96)00851-1}}.

\bibitem{centauro_bjorken}
J.~D. Bjorken, L.~D. McLerran, Explosive quark matter and the ''{Centauro}''
  event, {Phys. Rev. D} 20 (1979) 2353.

\bibitem{centauro_sqm}
S.~B. Shaulov, Evidences for strangelet presence in primary cosmic rays, {Acta
  Phys. Hung. N. Heavy Ion Phys.} 4 (1996) 403--422.
\newblock \href {http://dx.doi.org/10.1007/BF03155637}
  {\path{doi:10.1007/BF03155637}}.

\bibitem{pavlyuchenko}
V.~P. Pavlyuchenko, R.~M. Martirosov, N.~M. Nikolskaya, \etal, On the possibility of testing the models of the knee in {PCR}
  spectrum, {Bull. Lebedev Phys. Inst.} 41 (2014) 53--55.
\newblock \href {http://arxiv.org/abs/[astro-ph.HE/1406.0799]}
  {\path{arXiv:[astro-ph.HE/1406.0799]}}, \href
  {http://dx.doi.org/10.3103/S1068335614030014}
  {\path{doi:10.3103/S1068335614030014}}.

\bibitem{pavlyuchenko2}
V.~P. Pavlyuchenko, R.~M. Martirosov, N.~M. Nikolskaya, A.~D. Erlykin,
  Essential properties of the difference method for the search of the
  anisotropy of the primary cosmic radiation (2015).
\newblock \href {http://arxiv.org/abs/[astro-ph.HE/1509.09316]}
  {\path{arXiv:[astro-ph.HE/1509.09316]}}.

\bibitem{muons1}
R.~Ulrich, R.~Engel, M.~Unger, Hadronic multiparticle production at ultrahigh
  energies and extensive air showers, {Phys. Rev. D} 83 (2011) 054026.
\newblock \href {http://arxiv.org/abs/[hep-ph/1010.4310]}
  {\path{arXiv:[hep-ph/1010.4310]}}, \href
  {http://dx.doi.org/10.1103/PhysRevD.83.054026}
  {\path{doi:10.1103/PhysRevD.83.054026}}.

\bibitem{muons2}
K.~H. Kampert, M.~Unger, Measurements of the cosmic ray composition with air
  shower experiments, {Astropart. Phys.} 35 (2012) 660--678.
\newblock \href {http://dx.doi.org/10.1016/j.astropartphys.2012.02.004}
  {\path{doi:10.1016/j.astropartphys.2012.02.004}}.

\bibitem{muons3}
J.~Knapp, D.~Heck, S.~J. Sciutto, \etal, Extensive air shower
  simulations at the highest energies, {Astropart. Phys.} 19 (2003) 77--99.
\newblock \href {http://arxiv.org/abs/[astro-ph/0206414]}
  {\path{arXiv:[astro-ph/0206414]}}, \href
  {http://dx.doi.org/10.1016/S0927-6505(02)00187-1}
  {\path{doi:10.1016/S0927-6505(02)00187-1}}.

\bibitem{muons4}
T.~Pierog, K.~Werner, Muon production in extended air shower simulations,
  {Phys. Rev. Lett.} 101 (2008) 171101.
\newblock \href {http://dx.doi.org/10.1103/PhysRevLett.101.171101}
  {\path{doi:10.1103/PhysRevLett.101.171101}}.

\bibitem{muons-horizo}
M.~Ave, J.~A. Hinton, R.~A. V\'{a}zquez, \etal, A new approach
  to inferring the mass composition of cosmic rays, {Astropart. Phys.} 18
  (2003) 367--375.
\newblock \href {http://dx.doi.org/10.1016/S0927-6505(02)00158-5}
  {\path{doi:10.1016/S0927-6505(02)00158-5}}.

\bibitem{muons-horizo2}
{The Pierre Auger Collaboration}, Reconstruction of inclined air showers
  detected with the {Pierre Auger Observatory}, {J. Cosmol. Astropart. Phys.}
  2014 (2014) 019.
\newblock \href {http://arxiv.org/abs/[astro-ph.HE/1407.3214]}
  {\path{arXiv:[astro-ph.HE/1407.3214]}}, \href
  {http://dx.doi.org/10.1088/1475-7516/2014/08/019}
  {\path{doi:10.1088/1475-7516/2014/08/019}}.

\bibitem{ontienmuons1}
E.~V. Basarov, R.~U. Beisembaev, S.~P. Besshapov, \etal, The observation of muon groups at {Tien-Shan} station, in:
  Proceedings of the 14th {ICRC}, Vol.~6, M\"{u}nchen, Germany, 1975, pp.
  2067--2071.

\bibitem{ontienmuons2}
E.~V. Basarov, R.~U. Beisembaev, S.~P. Besshapov, \etal, The search for directly produced muon pairs in extensive air
  siowers, in: Proceedings of the 15th {ICRC}, Vol.~6, {Plovdiv, Bulgaria},
  1977, pp. 179--184.

\bibitem{ontienmuons3}
J.~N. Stamenov, N.~K. Georgiev, N.~V. Kabanova, \etal, Phenomenological characteristics of the muon component of
  extensive air showers at mountain level, {Nauka, Moscow}, 1979, pp. 132--151
  {(in Russian)}.

\bibitem{ontienhistory3}
I.~A. Amurina, V.~P. Antonova, G.~M. Autova, \etal, Current state of the
  {ATHLET} set-up at the {Tien-Shan}, {Nucl. Phys. B (Proc. Suppl.)} 151 (2006)
  422--424.

\bibitem{tieneng}
\href{{http://www.tien-shan.org}}{The {T}ien-{S}han {M}ountain {S}tation's
  {D}atabase}, {http://www.ti\-en-shan.\-org} (2006--2018).

\bibitem{shalour2006}
G.~I. Britvich, S.~K. Chernichenko, A.~P. Chubenko, \etal, The large scintillation charged
  particles detector of the {Tien-Shan} complex {'ATHLET'}, {Nucl. Instrum.
  Methods A} 564~(1) (2006) 225--234.
\newblock \href {http://dx.doi.org/10.1016/j.nima.2006.03.042}
  {\path{doi:10.1016/j.nima.2006.03.042}}.

\bibitem{smp04}
{Analog Devices}, \href{http://www.analog.com}{{CMOS Quad Sample-and-Hold
  Amplifier Datasheet}}, http://www.analog.com (2015).

\bibitem{ad7888}
{Analog Devices}, \href{http://www.analog.com}{2.7~{V} to 5.25~{V}, micropower,
  8-channel, 125~k{SPS}, 12-bit {ADC} in 16-lead {TSSOP} {AD7888} {Datasheet}},
  http://www.analog.com (2015).

\bibitem{nkg-greisen}
K.~Greisen, Progress in Cosmic Ray Physics, {North Holland Publ.}, Amsterdam,
  1956.

\bibitem{nkg-nk}
K.~Kamata, J.~Nishimura, The lateral and the angular structure functions of
  electron showers, {Prog. Theor. Phys. Supp.} 6 (1958) 93--155.

\bibitem{scipy}
{SciPy},
  \href{{http://docs.scipy.org/doc/scipy/reference/}}{{SciPy
  Reference Guide}}, {http://docs.\-sci\-py.\-org/doc/sci\-py/re\-fe\-ren\-ce/}
  (2015).

\bibitem{neldermead}
J.~A. Nelder, R.~Mead, A simplex method for function minimization, {Comput. J.}
  7~(4) (1965) 308--313.

\bibitem{aragats_fpr_b}
S.~Blokhin, N.~Kabanova, S.~Kazaryan, \etal, Lateral
  distribution function of {EAS} electrons in shower size range {$10^5\le N_e
  \le 3\cdot 10^7$} according to data of {Maket-ANI array}, {Bull. Russ. Acad.
  Sci. Phys.} 69 (2005) 535.

\bibitem{hadron_spc}
V.~V. Arabkin, L.~I. Vildanova, N.~G. Vildanov, \etal, Energy spectrum
  of primary cosmic particles in the energy range $10^{13}-10^{18}${eV}
  according to {Tien~Shan} installation data, {Bull. Russ. Acad. Sci. Phys.} 57
  (1993) 640--643.

\bibitem{aragats_spc}
A.~P. Garyaka, R.~M. Martirosov, S.~O. Sokhoyan, \etal, The
  study of the main characteristics of primary {VHE} cosmic radiation in
  {Gamma} experiment ({Aragats Mountain, Armenia}), {Bull. Natl. Acad. Sci.
  Armen., Phys.} 48~(2) (2013) 79--94.

\bibitem{ontienmonitor}
A.~G. Zusmanovich, O.~N. Kryakunova, A.~L. Shepetov, The {Tien-Shan} mountain
  cosmic ray station of the {I}onosphere {I}nstitute of {K}azakhstan
  {R}epublic, {Adv. Space Res.} 44~(10) (2009) 1194--1199.
\newblock \href {http://dx.doi.org/10.1016/j.asr.2008.11.030}
  {\path{doi:10.1016/j.asr.2008.11.030}}.

\bibitem{carmichel_supermonitor}
H.~Carmichael, C.~J. Hatton, Experimental investigation of the {NM-64} neutron
  monitor, {Can. J. Phys.} 42 (1964) 2443.
\newblock \href {http://dx.doi.org/10.1139/p64-222}
  {\path{doi:10.1139/p64-222}}.

\bibitem{simpson}
J.~A. Simpson, W.~Fonger, S.~B. Treiman, Cosmic radiation intensity-time
  variations and their origin. {I. N}eutron intensity variation method and
  meteorological factors, {Phys. Rev.} 90 (1953) 934.
\newblock \href {http://dx.doi.org/10.1103/PhysRev.90.934}
  {\path{doi:10.1103/PhysRev.90.934}}.

\bibitem{inca1998c}
K.~V. Aleksandrov, G.~T. Zatsepin, E.~P. Kuznetsov, \etal, On a possibility to increase the energy resolution and
  rejection capability of the {MINOS} and {NOE} neutrino calorimeters while
  detecting charge-current and neutral-current events, {Dokl. Phys.} 43~(12)
  (1998) 738--742.
\newblock \href {http://dx.doi.org/10.1088/0954-3899/28/2/306}
  {\path{doi:10.1088/0954-3899/28/2/306}}.

\bibitem{inca1999d}
{The INCA Collaboration}, {The INCA Project II}. {M}easurements of the neutron
  yield from a lead absorber for pion and proton projectiles, in: {Proceedings
  of the 26th ICRC}, Vol.~3, {Salt Lake City, USA}, 1999, pp. 195--198.

\bibitem{nmn2003}
A.~P. Chubenko, A.~L. Shepetov, V.~P. Antonova, \etal, Multiplicity spectrum of {NM64} neutron
  supermonitor and hadron energy spectrum at mountain level, in: {Proceedings
  of the 28th ICRC}, {Tsukuba, Japan}, 2003, pp. 789--792.

\bibitem{geant_base}
{Geant4 Collaboration}, {Geant4} -- a simulation toolkit, {Nucl. Instrum.
  Methods A} 506~(3) (2003) 250--303.
\newblock \href {http://dx.doi.org/10.1016/S0168-9002(03)01368-8}
  {\path{doi:10.1016/S0168-9002(03)01368-8}}.

\bibitem{hughes_supermonitor}
E.~B. Hughes, P.~L. Marsden, G.~Brooke, \etal, Neutron
  production by cosmic ray protons in lead, {Proc. Phys. Soc.} 83 (1964) 239.
\newblock \href {http://dx.doi.org/10.1088/0370-1328/83/2/308}
  {\path{doi:10.1088/0370-1328/83/2/308}}.

\bibitem{onmulti}
R.~A. Nobles, E.~B. Hughes, C.~J. Wolfson, Empirical response functions for a
  neutron multiplicity monitor, {Space Phys.} 74 (1969) 6459–6470.
\newblock \href {http://dx.doi.org/0.1029/JA074i026p06459}
  {\path{doi:0.1029/JA074i026p06459}}.

\bibitem{yanke2011}
A.~A. Abunin, E.~V. Pletnikov, A.~L. Shchepetov, V.~G. Yanke, Efficiency of
  detection for neutron detectors with different geometries, {Bull. Russ. Acad.
  Sci. Phys.} 75~(6) (2011) 866--868.

\bibitem{STM32F405}
{STMicroelectronics}, \href{http://www.st.com/}{{DM00037051} {STM32F405xx},
  {STM32F407xx} {D}atasheet -- production data}, http://www.\-st.\-com/ (2015).

\bibitem{inca2006}
V.~V. Ammosov, G.~I. Britvich, A.~P. Chubenko, \etal, The modern concept of the {INCA} project elements,
  {Nucl. Phys. B (Proc. Suppl.)} 151 (2006) 426--429.
\newblock \href {http://dx.doi.org/10.1016/j.nuclphysbps.2005.07.077}
  {\path{doi:10.1016/j.nuclphysbps.2005.07.077}}.

\bibitem{inca2007}
V.~V. Ammosov, G.~I. Britvich, A.~P. Soldatov, \etal, On potentialities of a multipurpose
  astrophysical orbital observatory in studies of high-energy cosmic rays,
  {Nucl. Phys. B (Proc. Suppl.)} 166 (2007) 140--144.

\bibitem{inca2008}
V.~Ammosov, G.~Britvich, A.~Soldatov, \etal, High-energy cosmic-ray
  physics study by multipurpose astrophysical orbital observatory, {Nucl. Phys.
  B (Proc. Suppl.)} 175 (2008) 190--193.

\bibitem{inca1999c}
{The INCA Collaboration}, {The INCA Project III}. {N}ew method for separation
  of electromagnetic and hadron cascades in detection of primary electrons and
  gamma-rays, in: {Proceedings of the 26th ICRC}, Vol.~3, {Salt Lake City,
  USA}, 1999, pp. 203--206.

\bibitem{inca2001}
K.~V. Alexandrov, M.~Ambrosio, V.~V. Ammosov, \etal, A new method of
  ionization-neutron calorimeter for direct investigation of high-energy
  electrons and primary nuclei of cosmic rays up to the knee region, {Nucl.
  Instrum. Methods A} 459~(1--2) (2001) 135--156.
\newblock \href {http://dx.doi.org/10.1016/S0168-9002(00)01006-8}
  {\path{doi:10.1016/S0168-9002(00)01006-8}}.

\bibitem{ontiencheren}
R.~U. Beisembaev, Y.~N. Vavilov, M.~I. Vildanova, \etal,
  The first results obtained with the installation {HORIZON-T}, {J. Phys.:
  Conf. Ser.} 409 (2013) 012127.
\newblock \href {http://dx.doi.org/10.1088/1742-6596/409/1/012127}
  {\path{doi:10.1088/1742-6596/409/1/012127}}.

\bibitem{space_weather_inbook}
H.~Mavromichalaki, V.~Yanke, L.~Dorman, \etal, Effects of Space Weather on Technology Infrastructure, {NATO
  Science Series II}, 2003, Ch. Neutron Monitor Network in Real Time and Space
  Weather, pp. 301--317.
\newblock \href {http://dx.doi.org/10.1007/1-4020-2754-0_16}
  {\path{doi:10.1007/1-4020-2754-0_16}}.

\bibitem{ongle}
A.~V. Belov, E.~A. Eroshenko, O.~N. Kryakunova, \etal, Ground
  level enhancements of solar cosmic rays during the last three solar cycles,
  {Geomagn. Aeron.} 50 (2010) 21--33.
\newblock \href {http://dx.doi.org/10.1134/S0016793210010032}
  {\path{doi:10.1134/S0016793210010032}}.

\bibitem{thunderourufn}
A.~V. Gurevich, A.~N. Karashtin, V.~A. Ryabov, \etal,
  Nonlinear phenomena in the ionospheric plasma. {E}ffects of cosmic rays and
  runaway breakdown on thunderstorm discharges, {Phys.-Uspekhi} 52~(7) (2009)
  735--745.
\newblock \href {http://dx.doi.org/10.3367/UFNe.0179.200907h.0779}
  {\path{doi:10.3367/UFNe.0179.200907h.0779}}.

\bibitem{rainsour2009}
N.~Salikhov, G.~Pak, O.~Kryakunova, A.~Chubenko, A.~Shepetov, An increase of
  the soft gamma-radiation background by precipitations, in: Proceedings of the
  32nd {ICRC}, Vol.~11, Beijing, China, 2011, pp. 368--371.

\bibitem{undgmouacou}
L.~I. Vil’danova, G.~A. Gusev, V.~V. Zhukov, \etal, The first results of observations of acoustic signals generated
  by cosmic ray muons in a seismically stressed medium, {Bull. Lebedev Phys.
  Inst.} 40~(3) (2013) 74--79.
\newblock \href {http://dx.doi.org/10.3103/S1068335613030019}
  {\path{doi:10.3103/S1068335613030019}}.


\end{thebibliography}

\newcommand {\etal}{et al. }


\end{document}